\documentclass[conference]{IEEEtran}
\IEEEoverridecommandlockouts

\usepackage[utf8]{inputenc}
\usepackage[T1]{fontenc}
\usepackage{graphicx}
\usepackage{amsmath}
\usepackage{amsthm}
\usepackage{amssymb}
\usepackage{units}
\usepackage{ifthen}
\usepackage{listings}
\usepackage{paralist}
\usepackage{xcolor}
\usepackage{xspace}
\usepackage{wrapfig}
\usepackage{float}
\usepackage[inline]{enumitem}
\usepackage{multirow, booktabs}

\graphicspath{{figures/}}

\usepackage[unicode=true,
  hidelinks,
  pdfauthor={Rafal Graczyk, Paulo Esteves-Verissimo and Marcus Voelp},
  pdftitle={Sanctuary lost: Cyber-Physical Warfare in Space.},
  pdfsubject={}, 
  pdfkeywords={},
  colorlinks=true,
  allcolors=black,
  citecolor=blue
]{hyperref}

\usepackage{cleveref}

\newboolean{showcomments}
\setboolean{showcomments}{false}

\ifthenelse{\boolean{showcomments}}
           { \newcommand{\mynote}[3]{
               \fbox{\bfseries\sffamily\scriptsize#1}
                    {\small$\blacktriangleright$\textsf{\emph{\color{#3}{#2}}}$\blacktriangleleft$}}}
           { \newcommand{\mynote}[3]{}}

\definecolor{orange}{RGB}{225,128,0}
\definecolor{brown}{RGB}{225,128,128}
\definecolor{purple}{rgb}{0.54, 0.17, 0.89}
\definecolor{rggold}{RGB}{218,165,32}
\definecolor{rgviolet}{RGB}{119,51,255}

\newcommand{\st}[1]{\mynote{ST}{#1}{red}} 

\newcommand{\rg}[1]{\mynote{RG}{#1}{rgviolet}}

\ifthenelse{\boolean{showcomments}}
           { 
           }{}

\definecolor{darkgreen}{RGB}{0,155,0}
\begin{document}

\title{Sanctuary lost: a cyber-physical warfare in space} 

\author{
  \IEEEauthorblockN{
    Rafal Graczyk\IEEEauthorrefmark{1},
    Paulo Esteves-Verissimo\IEEEauthorrefmark{2}, and 
    Marcus V\"olp\IEEEauthorrefmark{1}}
  \IEEEauthorblockA{
    \IEEEauthorrefmark{1}University of Luxembourg, 
    Interdisciplinary Center for Security, Reliability and Trust (SnT) - CritiX group\\
    \IEEEauthorrefmark{2}King Abdullah University of Science and Technology (KAUST), 
    Resilient Computing and Cybersecurity Center (RC3)\\
    \IEEEauthorrefmark{1}\{rafal.graczyk, marcus.voelp\}@uni.lu
    \IEEEauthorrefmark{2}paulo.verissimo@kaust.edu.sa
  }
}

\maketitle

\begin{abstract}
  Over the last decades, space has grown from a purely scientific
  struggle, fueled by the desire to demonstrate superiority of one
  regime over the other, to an anchor point of the economies of
  essentially all developed countries. Many businesses depend
  crucially on satellite communication or data acquisition, not only
  for defense purposes, but increasingly also for day-to-day
  applications. However, although so far space faring nations
  refrained from extending their earth-bound conflicts into space,
  this critical infrastructure is not as invulnerable as common
  knowledge suggests. In this paper, we analyze the threats space
  vehicles are exposed to and what must change to mitigate them. In
  particular, we shall focus on cyber threats, which may well be
  mounted by small countries and terrorist organisations, whose
  incentives do not necessarily include sustainability of the space
  domain and who may not be susceptible to the threat of mutual
  retaliation on the ground. We survey incidents, highlight threats and
  raise awareness from general perparedness for accidental faults,
  which is already widely spread within the space community, to
  preparedness and tolerance of both accidental and malicious faults
  (such as targeted attacks by cyber terrorists and nation-state hackers).
\end{abstract}

\begin{IEEEkeywords}
  space, satellite, cyberphysical, system, threat vector,
  cybersafety, cybersecurity
\end{IEEEkeywords}

\section{Introduction}
\label{sec:introduction}

Space infrastructure in itself is not a very large business: with 366
billion USD of global revenues \cite{Bryce20a}, it constitutes
$\sim$0.42\% of the global economy of 87.5 trillion USD in 2019
\cite{WorldGDP, WorldGDPwiki}.

Space assets used to be of military origin, since the 80's of XX century, are
increasingly civilian, starting from telecommunications, then spreading to other
fields of near space exploitation. Today, many other sectors of our economy,
even if, apparently ground-based, crucially depend on the space infrastructure,
including \cite{Hollingham13}: mass media, global transport and logistics,
military, intelligence, utilities, agriculture, banking, oil and mining.
Activities from private companies such as SpaceX with Starlink and
OneWeb complement this list with megaconstellations providing
communication and fast internet to remote locations.

This dependence turns space into an important resource, in particular
for the developed countries and their economies, space faring and not, and
a worthwhile target for protection.
Unfortunately several threats put the sustainable use of space at
risk, both as a foundation for military operations, but more
importantly for the economic applications that affect our daily life.

In this paper, we review past incidents and derive a comprehensive
threat-plane for space and the orbits therein as sustainable resource
to deploy satellites for a multitude of purposes, but also for human
and robotic scientific missions. We consider both threats from
accidental causes, but more importantly also threads originating from
intentionally malicious activities, such as targeted cyber and/or
physical attacks.

As we shall see, threats do not only originate from other space-faring
nations and their ability to damage, shoot down or disrupt
communication of other space-faring nation's vehicles. They may also
originate from terrorist groups and the nation-state hackers of less
developed countries who would not suffer from disrupting access to
space for the upcoming decades until effective debris-removal
technologies become available. In particular, not all of these groups
are vulnerable to the common threat of retaliation on the ground, which
several space faring nations formulated as balance of powers, should
space vehicles of that nation be attacked. Even the correct
attribution of attacks will be difficult, as can be seen in some of
the recent cyber attacks that keep happening on ground.

We assume this topic may as well be interesting for a lay audience,
not necessarily proficient in the details of space flight or in the
way cyber attacks can manifest. In addition to our threat-plane
analysis, we shall therefore also introduce the background necessary
for understanding these aspects, inviting the proficient reader to
skip these sections. Our contributions include in
Section~\ref{sec:infrastructure} an analysis why space forms a
critical information infrastructure worth protecting, a taxonomy of
kinetic and non-kinetic threats space vehicles are exposed to (in
Section~\ref{sec:threat-classifications-general}), which we shall
apply in our survey of past incidents during the various stages of
space vehicle development, deployment and operation (in
Section~\ref{sec:attacks}). Where appropriate, we suggest measures
which may help overcome some of these threats, however ultimately we
recommend leveraging on the care already taken to tolerate, survive
and repair the consequences of accidental faults and to extend this
care to also become resilient to intentionally malicious faults. We
begin by surveying space flight and the foundations on which it is
based.

\section{Space Flight 101}
\label{sec:overview}
\st{ I assume lay audience might lack either aerospace or CS / cyber background
 so I try to bridge those 2 domain, and also show wider context of the the
 topic: What are space systems and why they are slipping out of out intuition,
 I start with introduction to the field from aerospace side, where we are and
 how it changes in coming years, megaconstellations, space economic activity,
 the dangers of sustainability}

Any projectile (ball, rock, axe) thrown at ground level, tangential to a
planet's surface, will closer or farther fall down. That's what everyday
experience tells us and this intuition is generally true, unless the projectile
is thrown so fast that it falls at the same rate, that the ground recedes. The
Earth is a geoid. However, for our little thought experiment we approximate it to
be spherical, receding roughly 5 meters for every 8 km traveled along the
tangent. The exact speed, required for maintaining the hypothetical trajectory,
following planet's surface, depends on the planet's mass and is called first
cosmic velocity. For Earth, this velocity is roughly 7.9~km/s.

Changing in our experimental setup the projectile to a satellite, our
throwing hand to a rocket and raising the target altitude to above the
atmosphere, where there are no buildings, hills, weather and last, but
most importantly, no atmospheric drag, and we have entered the space
domain and successfully placed our satellite on a path called the orbit.

Raising the orbit's altitude, we reduce the velocity required to maintain a
circular orbit (one with the same altitude above Earth at all times). At the
same time, we increase the orbital period, which at low Earth orbits (LEOs) is
about 90 minutes. Orbital period is the time after which a satellite returns to
the same position above ground, here after 1.5 hours, not yet compensating for
the rotation of the Earth.

Observers on the ground will still notice the satellite several hundreds
if not thousands of kilometers west, although the spacecraft returned
to its initial position. This is because our planet rotates at a
linear velocity of 1670 km/h at the equator, which decreases as the
latitude increases (e.g., to 1180 km/h at 45 degrees and to 0 km/h at
the Poles, at 90 degrees).
For satellite operators this implies limited 
visibility of the satellite and hence also limited communication
possibilities.  Sometimes, depending on the exact location of the
ground station relative to the satellite's orbit, convenient revisits
may happen only after days.
%
%

A satellite's orbiting plane can be oriented in all possible angles
with respect to the Earth's rotation plane. Orbital planes can be
equatorial, so that satellites will travel over the equator, or they
can be more inclined, allowing satellites to visit (observe/
communicate with) higher latitudes, up to polar orbits where satellites
will fly over each of the Earth's poles.

As mentioned before, the higher the altitude of a satellite's orbit, the longer
its orbital period. At roughly 36,000 km above Earth (about 10 \% of the
distance to the Moon), the orbital period becomes almost 24 hours, the same
length as Earth's rotation period.  Satellite on such an orbit are
geosynchronous and will visit each place on its path at exactly the same time
of day. If geosynchronous orbits have an inclination of zero degrees with
respect to the equator, the orbit becomes geostationary and satellites on this
orbit stand still over the same place on Earth, as they have the same angular
velocity as the Earth rotating below them. As this happens 36000 km above the
Earth's surface, the whole planet's hemisphere can be in view, which is a neat
feature for telecommunication or observation purposes, with obvious drawbacks
caused by the distance from ground that these satellites must have: the
propagation of radio signals causes noticeable latency and very sophisticated
optical and attitude stabilization systems are needed to maintain accurate
satellite pointing. Typical orbits and their applications are summarized in
Table~\ref{tab:orbits-summary}.


\begin{table*}[h!]
  \begin{center}
    \begin{tabular}{ p{0.2\textwidth} | p{0.2\textwidth} | p{0.5\textwidth} }
      \toprule
      Type & Altitude [km] & Remarks \\
      \midrule
      VLEO Very Low Earth Orbit & 180 - 500 & Earth observation (EO), very low communication latency; high imaging resolution with mildly sophisticated equipment; atmospheric drag causes the satellite to require propulsion to maintain the altitude; operation altitude of the International Space Station (ISS) \\ 
      LEO Low Earth Orbit & 500 - 2000 & EO, low communication latency; popular for microsatellites \\
      SSO Sun-Synchronous Orbit & same as LEO & subtype of LEO; low, retrograde orbit, sun synchronous; revisits the same spot at the same tame of day (the same shadows cast by objects), used for observation satellites; if traveling along the terminator line (dawn-dusk orbit) satellite could spent majority of time in sunlit conditions --- a preferable arrangement for power constrained satellites\\
      MEO Medium Earth Orbit & 2000 - 35786 & used by Global Navigation Satellite Systems (GNSS) systems, some telecommunication constellations(i.e. SES O3b)\\
      HEO Highly Elliptical Orbit & perigee and apogee vary & highly elliptical orbit, used for communication satellites or science equipment\\
      GTO Geo Transfer Orbit & perigee and apogee vary & subtype of HEO; used for accessing GEO and beyond (cis-lunar space, lunar orbits and deeper into the Solar System)\\
      GEO Geosynchronous Orbit & 35786 & telecommunication and observation satellites, if over equator, is Geostationary Orbit\\
      GEO Graveyard & $\sim$36300 & decommissioned GEO satellites shall end up here \\
      \midrule
    \end{tabular}
    \caption{Summary of most popular orbits and their applications \cite{Fritz2013, SSI19, Horowitz15, Reesman20}}
    \label{tab:orbits-summary}
  \end{center}
\end{table*}

\subsection{Fuel and Power Considerations}
\label{ssec:fuel-constraints}
\st{what challenges do space systems face, platform, payload division,
  segments}

Satellites above the atmosphere operate in vacuum, which very quickly
becomes very deep and provides so little residual atmospheric drag
that for precise satellite modeling purposes, other factors begin to
dominate, such as the influence of solar radiation.

Satellites and probes do not necessarily need propulsion to stay on orbit or to
maneuver and change attitude. This is obviously a bit of a too idealistic take
on orbital mechanics, as non-homogeneity of the gravitation field, influence of
other celestial bodies, solar wind and pressure, changes of the residual
atmospheric drag (e.g., due to high Sun activity), etc., perturb the stability
of the orbit, both in its plane arrangement and in its altitude.
In most parts however, it can be assumed that bodies injected into a
stable orbit will stay on this orbit and that changing the
orbit requires some kind of propulsion, such as rocket engines, cold
gas thrusters, resitojet thrusters, and various kinds of electric
engines (e.g., ion or Hall effect).

While particular implementations may vary significantly, generation of thrust
requires mass to be ejected for the spacecraft to experience a
force in the opposite direction. The bigger the mass times its
ejection velocity, the bigger will be the force that acts upon the
spacecraft and also the acceleration that the spacecraft experiences in the 
direction of this force.

Stored fuel and the spacecraft's propulsion capacity to change it's
velocity thus define how fast the craft can change its orbit and
ultimately constrains its mission and the maneuvers it can execute
during that time.

Some changes of the orbit require only reasonable amounts of fuel, especially
raising and lowering the perigee or the apogee of an orbit(i.e., the point
closest to / farthest from the Earth's center). Both points can be adjusted
using Hohmann maneuvers, which require engine burns tangential to the orbit in
one of these points to affect the opposite one (for details refer to 
\cite{Vallado13, Kluever18, Dekoulis18, Reesman20}). Other maneuvers are more
demanding and thus affect mission time more severely. For example, changing the
inclination of a satellite requires redirecting its momentum, most of which has
been built up at the launch and with the launch vehicle's propulsion. Not many
of these maneuvers can be performed with the limited fuel and propulsion
capacity of the satellite. Spacecraft injected by its launch vehicle or by its
apogee motor into an orbit are therefore typically left in this orbit. Apogee
motors are typically found in GEO satellites to make the orbit circular. Space
plane concept, which theoretically, could help to overcome satellite
maneuverability limitations, at the time of writing of this paper, is very
immature technology in practice \cite{X37B, ChinaSpaceplane}.

Spacecraft are also very constrained on the side of (electrical) power supply.
In practice, there are two primary power sources - solar arrays (suitable for
use in inner Solar System, beyond Mars, up to asteroid belt) and radioisotope
thermoelectric generator (RTG). Solar arrays take sunlight and convert (up to
30-35\% of it) into electric power. Solar arrays in order to reach higher power
output need to occupy large areas which requires them to be deployable
(they have to reliably unfold), which in turn makes the attitude control more
challenging. RTG's require heatsinks which are dead mass which increases the
launch cost, not to mention, that application of this technology is limited by
environmental and political constraints~\cite{Larson92, Wertz11}.

To sum up, both, electrical power, for supplying the spacecraft, and fuel, for
managing the spacecraft orbit, are ultimate currencies in which each spacecraft
designer and operator has to pay for owned vessel functions and capabilities.
Since the spacecraft design and operation concepts are the results of a long and
tedious process of architectural trade-offs it is usually very hard to add new
function to already operating systems or to maneuver them in way that was not
foreseen in the first place. Since placement in orbit, spacecraft is bound to
follow it's preconceived destiny and the margins that could be used for
changing mission objectives are minimal, and, typically, are left for emergency
or fault contingency operations. 



\subsection{On Board Systems}
\label{ssec:on-board-systems}
Space systems operate in the harsh environment of outer space,
remotely and, during times when no direct connection with ground
stations is possible, with limited supervision. 
In addition to operating the instruments required to fulfill the satellites
mission (sensors, cameras, communication relays, etc.), which are called the
payload part, the satellite must also secure its own survival over the
envisioned mission time and beyond. Despite shorter initially foreseen
lifetimes (e.g., of a few months on low orbit, a few years for regular
satellites or 15 years for GEO), due to limitations outlined in 
Section~\ref{ssec:fuel-constraints}, satellite lifetimes can go far beyond this initial
plan (e.g., 20 years and counting for the ISS~\cite{ISS} and 43 years and
counting for the two of Voyager spacecraft~\cite{Voyager}).

The platform part assumes this role of securing a satellite
survival. More precisely, it ensures sufficient resources remain
available for the payload systems to fulfill their mission. The
platform also controls power \& thermal management, propulsion and
thus orbit and attitude control as well as the command and telemetry
links \rg{references}.

Depending on the type of mission, payloads might require more or less
supervision, may have own communication links (both, up- and
down-links), but are otherwise isolated from the platform part. The
platform supplies the payload system with resources, commands and
receives telemetry in return. In all other aspect, the platform
remains separate from the payload subsystem, as its ultimate role is
to ensure the operational safety of the satellite as a whole.

\subsection{Space Systems beyond Space}
\label{ssec:space-systems-beyond-space}

Space operations consist not only of the vehicles deployed in space
(satellites, stations, spacecraft), but also include systems on the ground for
operation planning, control, tracking and communication, as well as the launch
vehicles required to deploy spacecraft in space. The segments are called space,
ground and launch segment, respectively. 
Ground segment, consists mainly of ground stations (used to establish
communication with spacecraft platform, payload or both) and mission control
facilities where commanding of spacecraft takes places. Tracking facilities,
used to be assumed as a part of ground segment, but nowadays, due to
significant increase in space congestion, traffic management becomes recognized
as another, independent, segment of space system architecture. Here, we shall
return to specifics of the ground-space interactions outlined at the beginning
of this chapter, and remind that any ground based facility is limited in it's
capability to establish communication with (or track) the satellite. Likewise,
satellite is able to interact (i.e. provide the service) with entities withing
it's field of view, during the time of pass (with the obvious advantage of
geostationary satellites having constant field of view and capacity to provide a
continuous service).
Hence, the ground-space interaction limitations impose that:
\begin{enumerate}
  \item With a few exceptions, ground stations lose contact once the
  satellite leaves their observation cone. Unless the provider has
  access to geographically spread ground stations to continuously
  track the satellite, the vehicle must operate autonomously; and
  \item Having observed the satellite, adversaries may predict when it
  comes into range of a critical operation to conceal their activity
  or to take countermeasures against the satellite.
\end{enumerate}
In addition, there is so called user segment, which consists of user terminals
for telecommunication services~\cite{VSAT} but also satellite navigation
receivers~\cite{navigation-receiver}, AIS~\cite{AIS} or 
ADS-B~\cite{ADS-B} transmitters or other, similar, systems. Organizations, facilities and
infrastructure that is used to process or distribute acquired information to
interested parties are also considered as a part of user segment.

\subsection{Crewed vs Robotic}
So far we discussed mainly about satellites, but spacecraft can be capable of
supporting human space flight. Main differences between crewed and robotic
activities, are the stakes (loss of life vs loss of equipment), complexity
(life support systems and operations), capabilities (crewed missions are much
more flexible as utilize human invention) and purpose (robotic are there to
build infrastructure for services or explore deep space, while crewed are for
science, outreach). While crewed operations are, for the time being, limited to
few stations orbiting the Earth on LEO~\cite{ISS, Tiangong}, this is expected
to change in the near feature with human expansion to cis-lunar space~\cite{Artemis},
 following the scientific steps of Apollo program~\cite{Apollo} and
building foundations for sustained human presence beyond Earth and extending
technical and economic activity to other celestial bodies. Today, crewed
missions are primarily limited to science missions and sophisticated repairs
where human flexibility to react to unforeseen situations outweighs the
additional weight and complexity of life-support systems and the higher risks
of the mission. Whereas loss of equipment is an unfortunate but tolerable
event, loss of life remains unacceptable.


\subsection{Life cycle}
\label{ssec:life-cycle}
\st{long lifecycle: when build according to space related standards and qualified to be used in space, or having as successful space operation heritage drives the system vendors to reuse the platforms as long as mission payload requirements and user operational constraints are met. Thus the designs and hardware of a systems deployed nowadays may have been conceived one or two decades ago, with quite different mindsets and thereat perceptions.}
A peculiar aspect of space systems is the length of their life cycle and
associated costs. The time from conception and preliminary design to the
disposal of the first unit of a future space system (be it a satellite or a
spacecraft) can easily span 2 or more decades. The last units remaining in
operation can be decommissioned after 3-4 decades since the time they were
designed.\\


This has several reasons:
First, space projects are managed by large governmental agencies which aim to
minimize project organizational and technical risks by enforcing bureaucratic
processes and adherence to space, industrial, standards~\cite{ECSS, NTSS} which
typically require extensive documentation of the work done and planned.
While justified for crewed missions or high profile missions, these processes
are not always necessary for shorter, cheaper, experimental missions, which is
the trend already visible in the community and is described in greater details
in following section~\ref{ssec:new-space}.
Stretching the design process in time increases the costs of a
mission, in particular in terms of the staff required to follow these
processes.

Second, design processes are limited to equipment that is built exclusively from
qualified components, which are carefully tested for their suitability for
space. Component qualification involves a significant number of tests, many of
which lead to the destruction of the units under test, and require careful
documentation. Again labor intensive tasks have to be conducted for the purpose
of providing assurance to customers or stakeholders.

Third, these costs and manpower needs are further magnified by several levels of
subcontracting and can easily lead to a price factor of 10--100 compared to
commercial of the shelf (COTS) equipment or component that is not space
graded.

A fourth aspect worth considering in the space segment life cycle, is
that changes to already established and qualified designs nullifies
the qualification status of this design and requires repeating the
above processes. Shortcuts in the form of so called
delta-qualifications are only possible if the change is small and
still requires re-testing of all critical aspects related to that
change.

This has two consequences:
\begin{enumerate}
\item Due to the high costs of the qualification process, new
  equipment gets subjected to it only if the additional performance
  justifies such expenses; and
\item Due to the time needed, space graded equipment often lags
  10--15, or more, years behind COTS equipment.
\end{enumerate}

We shall return to this when reflecting about the threats to which
spacecrafts are exposed.



\subsection{New Space}
\label{ssec:new-space}
A careful reader will by now already have spotted vulnerabilities due
to inefficiencies in the classical way of development and deployment
of space vehicles and the infrastructure they need on the ground. Let us
therefore also introduce a recent trend towards a more lean process
for building space infrastructure, whose risks and threats we will
analyze as well: \emph{NewSpace}~\cite{Satsearch-NewSpace}.

The NewSpace movement gains popularity due to its promise of more affordable
satellites and launchers.  NewSpace aims to build upon COTS technologies and
components that, if qualified at all, undergo a much more relaxed testing
regime~\cite{ECSS-tailoring, Cubesat101}, accepting higher risks of failure for
the sake of improved performance. The promise of NewSpace applies in particular
to LEO, where harsh environmental influence is present, but weaker, than in
higher orbits.

%
While by no doubt, the trend is clear and hundreds of new companies have already
entered this market, which, as of 2020, grew to 26.2~BUSD invested in space
start-ups since 2015 (at the scale of 36.7~BUSD invested since 2000)~\cite{Bryce21a}.
That is truly enormous progress, but there are also few downsides
of such fast paced development process--the number of small satellites
(from sub-kg to 600 kg of mass) launched every year, since 2017, is more than
three hundred~\cite{Bryce20b} and is increasing, from 2010 to 2020, number of
actively operating satellites has grown 252\%~\cite{Bryce21c}.

NewSpace in it's dynamics and consequences is a process of rapid
economic expansion and unbounded exploitation of resources, very
similar to examples on Earth, which have lead to overexploitation and
environmental devastation. Outer space, especially LEO, used to be
free of debris, but is now populated by a large number of new
spacecrafts entering into service, sometimes even from a single
launch, with debris (primarily from upper stage operations) starting
to accumulate beyond tolerable dimensions.

Increasing orbital congestion combined with a general disregard of
long term sustainability of the space environment (i.e. failure to
ensure fast deorbiting when satellites are disposed) are first steps
towards Kessler's syndrome~\cite{Kessler78} and we still have no
capacity to purge orbits from dead satellites and debris.

Reduced costs also originate from higher turnaround times of a few month between
contract signature and orbit injection rather than years. The lower costs and
simplified quality assurance results in, higher, than in Old Space, mission
failure rates. For example, Jacklin~\cite{Jacklin19} found that around 40\% of
all small satellite mission launched between 2009 and 2016 failed partially or
fully. It is also worth mentioning that Jacklin excluded academic endeavors
from this analysis as their success if often defined very broadly. It is also
fair to indicate that further industrialization of processes in NewSpace domain
will reduce expected failure rates, as it is already seen with Starlink,
trending towards 3\% failure rate~\cite{physorg-starlink}. However, special
care must be exercised to not let the same time-to-market mechanisms that are
also present in other weakly regulated fields cause security to be treated as
an aftermath.


NewSpace is on course of enabling satellites to become interconnected, creating
orbital networks with many nodes and numerous points of entry that are
eventually connected to the Internet. It leads to the creation of
(mega-)constellations (i.e., formations of spacecrafts cooperating in achieving
a common goal, typically for telecommunication but also for real-time Earth
observation and similar activities~\cite{megaconstellation}), which on the one
hand enables operators and users to utilize the greater potential of these new
services and increases the availability and robustness against accidental
faults.
At the same time, NewSpace approach, and use of large constellations in
particular, also increases the attack surface, making it harder to defend and
maintain control on the system. The trend of increasing the size of satellite
constellation along with simplifying and miniaturizing the satellites
themselves starts to spill into the traditional space industry~\cite{O3b_mPOWER},
and most likely will become even more significant in the future.

\subsection{Space sustainability}
\label{ssec:sustainability}
\begin{figure}
  \includegraphics[width=\columnwidth]{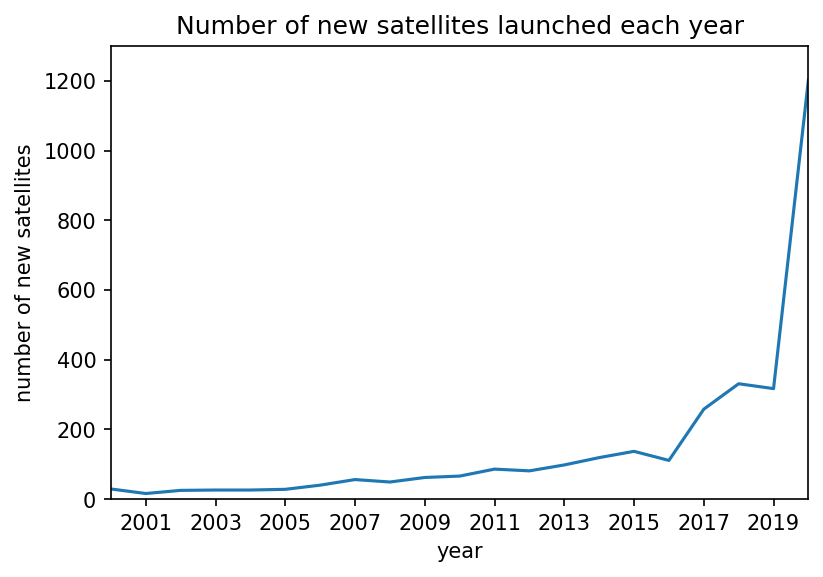}
  \caption{Number of new satellites launched each years. Data source:~\cite{UCS21}}
  \label{fig:obj-count}
\end{figure}

\begin{figure}
  \includegraphics[width=\columnwidth]{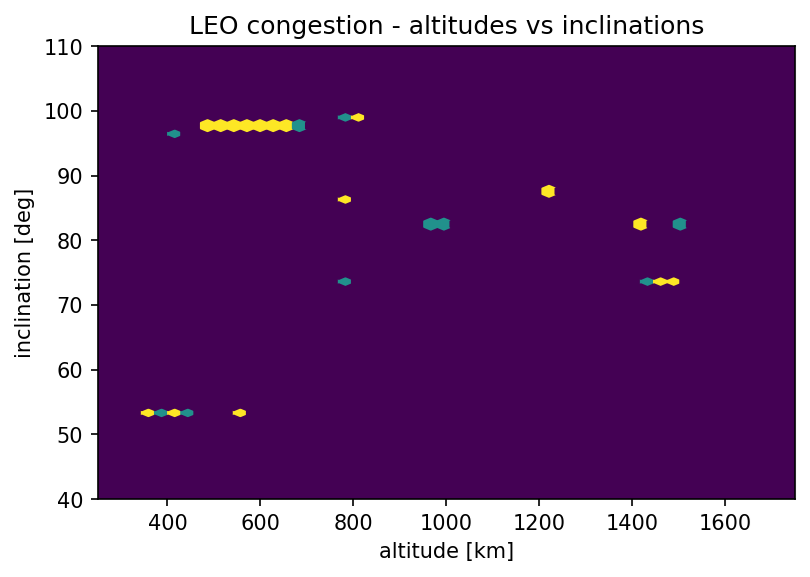}
  \caption{Congestion zones in LEO. Data source:~\cite{space-track}}
  \label{fig:congestion-zones}
\end{figure}

In Section~\ref{ssec:new-space} we have already mentioned that space near Earth,
especially LEO, is exploited beyond it's capacity to naturally clean itself
from dead spacecraft and launch and deployment debris (not to mention fields of
debris resulting from collisions). The Kessler syndrome~\cite{Kessler78} is an
existential threat to humankind capability to use space as part economy and
explore space as a part of our destiny. In 2021, caused, by both extensive
utilization and debris creation, we are able to track above 20 thousands objects in
near-Earth space. In Figure~\ref{fig:obj-count} it is also visible we're
already on exponential curve of growth of number of new satellites~\cite{UCS21}. What is even
worse, despite vastness of space, the object are placed on orbits which are
convenient or preferable for a given kind of activity. By this, zones of heavy
congestion are formed~\cite{ESA21env}. The described process increases the
catastrophic failure risks not only for regular LEO operations but also all
mission that in order to be deployed in deeper space, have to pass through
heavily congested and, debris filled, regions (refer to Figure~\ref
{fig:congestion-zones}). While Kessler syndrome was conceived as accidental
phenomenon arising purely from orbital mechanics and statistics, it can be
negatively augmented by malicious activity of competing space powers.
Anti-satellite weapon tests have already contributed significantly to amounts of
debris humanity has to deal with~\cite{ASAT-debris-fallout}.\\
What if collisions in space start becoming a result of intentional weaponization
of the satellites?

\subsection{Understanding the space domain}
\label{ssec:space-domain}
\st{summing up what I wish reader have realized so far}.\\
By now, it shall be apparent to the reader that systems of the space segment
cannot be analyzed without taking into consideration the environment in which
they operate. As famous Robert Hainlein is often quoted: "Once you’re in orbit,
you’re halfway to anywhere". The reign of gravity can't be overcome, future
satellite's positions can be only tweaked and will remain to large extent
predictable. In space operations, counter-intuitively, physical closeness is
not a measure of capability to link and physical separation is not a measure of
capability to interact~\cite{Oberg99}.
The physical environment, orbital considerations, Sun activity, the Earth's
magnetosphere and the electromagnetic spectrum defines all space segment
capabilities and limitations.
The moment a satellite system is injected into its orbit and set up for
operation, the margins for corrections and repairs have become very
thin and are often limited to what the satellite provides in the first
place.
In particular after the end of the Space Shuttle area, human repair
capabilities are severely limited and thus limited to robotic refurbishing
missions. 
Consequently, all malfunctions that have not been taken into account
during the design process, put the mission at risk.
While the space engineering community has large appreciation for and prepares
for tolerating accidental faults, there are still large gaps in understanding
the full consequences of malicious behavior. Accidental faults follow well
known statistics. In contrast, the drivers behind intentionally malicious
faults, such as targeted attacks, remain the intention and incentives under
which adversaries operate, not the statistics.
The gap in understanding the threats to space systems and scale of societal and
military dependence on space technologies and equipment, turns space into a
critical infrastructure, which we analyze next in Section~\ref
{sec:infrastructure}.


\section{Space is a Critical Information Infrastructure}
\label{sec:infrastructure}

\st{ There is a huge dependency on space infrastructure, in most cases it is not
 even well understood. Fragility of space infrastructure is underestimated not
 to mention preparedness for the possibility of loosing all those capabilities,
 and providing some back-up / resilience }

In Section~\ref{sec:introduction} we have already seen that despite space
infrastructure in itself not being a large business, large parts of
our economic wealth and growth depend on it.

Global economy thrives thanks to safe and efficient navigation on the high seas
provided by GNSS systems, with the help of AIS~\cite
{MarineTraffic-portal}. Aircraft, which move people and goods on large
distances, also depend on uninterrupted GNSS system operations, supported by
ADS-B reporting~\cite{FlightRadar-portal}. Oil rigs and pipelines, report the
telemetry via communication satellites ensuring remote management of
production~\cite{Burke96}. Some mines are already using autonomous equipment
which relies on both localization and communication capabilities provided from
space~\cite{auto-haulage}.

If any of the the space dependent services becomes disrupted, chaos ensues.
Without telemetry reports the whole production facilities have to stop to
prevent infrastructure damage. All the vessels that require to be localized and
navigated, have to stop to prevent crashes. It is not that there are no
alternatives - they are, but it will take time to deploy them in the safe manner.
The economy will stop for some time.

\begin{figure*}
  \begin{center}
    \includegraphics[width=.6\textwidth]{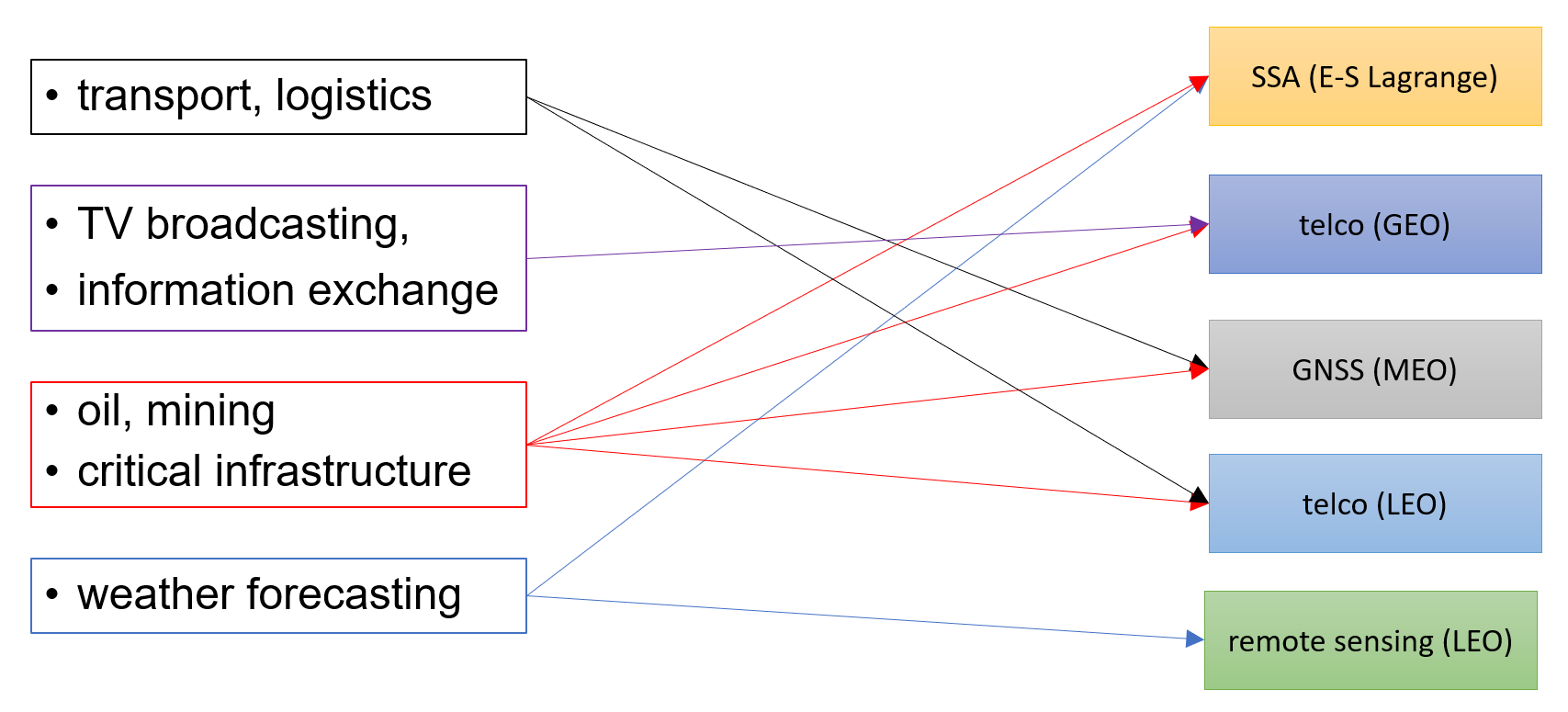}
    \caption{Global economy segments dependency on space infrastructure}
    \label{fig:economy-dep-space}
  \end{center}
\end{figure*}

From the above example and the examples given in Section~\ref{sec:introduction},
it is quite safe to write that most of our current and near
future modern economy is in one way or another dependent on uninterrupted
access to space as a conduit for information extraction and/or exchange.
Crippling space infrastructure therefore means interfering with these sectors
or, in less optimistic scenarios, disabling large parts of a country's economy
(see Figure~\ref{fig:economy-dep-space} for an overview of the dependence of
individual economy segments). It should be noted that this dependence is
particularly pronounced in developed countries, which justifies some of the
adversary models we shall consider.

The danger of the current situation is that, both the degree of
dependency nations have on space infrastructure and the depth of
susceptibility of that infrastructure, are widely underestimated. Huge
risks are looming from events (accidental or intentionally malicious
ones) that may cripple a nation's access to space and thereby its
economy and in part also its defense capabilities. Very little is done
to increase the robustness and resilience of the space systems we rely
on. The risk of a "Space Pearl Harbor" has been identified and
announced as early as in 2001 and now applies to a majority of developed
countries~\cite{Stoullig01, Spring05}.

This lack of resilience is partially caused by a phenomena called the
"High Ground Fallacy", a concept originating from military doctrine of
the old that armies would gain an advantage of being higher up and
therefore unreachable by their enemy.
%
%
Today, space is often called the new high ground, even in a civilian
understanding of space systems. However, space assets are not the fortresses to
seek shelter from the enemy or to descent on unprepared rivals. Instead, they
are among the most fragile, remote, outposts that support main forces with
information~\cite{Oberg99}. They are well reachable, both in the cyber-space
and physically, as we shall see in the remainder of this survey. In particular,
we shall see that they are reachable by entities insusceptible to the threats
of space access denial or retaliation, which turns protecting space systems
against advanced and persistent threats, the only viable option.



\section{A Taxonomy of Space Threats}
\label{sec:threat-classifications-general}

Space systems are exposed to a variety of threats and there are
several ways to classify the latter. Previous analyses study threats
to space systems from the perspective of international affairs,
intelligence or military organizations, which tends to introduce bias
of threats and countermeasures towards competing countries, namely
other space powers.
%
%
Threats investigated in that light are usually divided into two types:
kinetic and non-kinetic, pertaining to type of energy exchanged
between target and effector. An example threat classifications, which
follows the above principle, is summed up below:

\begin{itemize}
  \item {\bf kinetic threats}
  \begin{itemize}
    \item ground station physical attack \cite{Harrison21}
    \item direct ascent anti-satellite weapons \cite{Harrison21}
    \item co-orbital attack system, capable of launching projectiles \cite{Harrison21}
    \item collision with other satellites (both active and defunct), debris, upper stages
  \end{itemize}
  \item {\bf non-kinetic threats}
  \begin{itemize}
    \item high altitude nuclear detonation \cite{Harrison21}
    \item high-energy directed weapons \cite{Harrison21}
    \begin{itemize}
      \item lasers \cite{Harrison21}
      \item microwaves \cite{Harrison21}
    \end{itemize}
    \item electronic 
      \begin{itemize}
        \item uplink jamming \cite{Harrison21}
        \item downlink jamming \cite{Harrison21}
        \item spoofing \cite{Harrison21}
      \end{itemize}
    \item cyber
      \begin{itemize}
        \item data intercept / monitoring \cite{Harrison21}
        \item data corruption \cite{Harrison21}
        \item seizure of control \cite{Harrison21}
      \end{itemize}
  \end{itemize}
\end{itemize}

Threat classifications like the above, mainly due to the interests and
focus of the organizations that published them, discuss threats in
relation to global space powers (USA, Russia, China, India, perhaps,
Japan, France, Israel, for some, also Iran and North Korea,) that may
implement them (compare for example~\cite{Horowitz15, UNIDIR3,
  UNIDIR5, SSI19, Harrison21, Weeden21}).
The assumption is that exercising these threats requires capabilities
of limited availability (typically only to nation states), including
sophisticated launcher technologies, state-of-the art navigation and
tracking (for kinetic attacks, but also, for ground based laser
attacks or co-orbital microwave attacks), robotic rendezvous and
proximity technologies (for orbital operations like the deployment of
inspector spacecrafts with intended missions such as refueling,
maintenance or operational life extension, covering activities up to
Geostationary Orbits, but which can also be diverted from such
objectives to bring down space vehicles), and outstanding, close and
far range GNSS signal technology (e.g., for spoofing attacks). Table
\ref{tab:classic-threats-examples} summarizes most notable
nation-state space powers and their counterspace capabilities, while
\cite{SSI19, Harrison21, Weeden21} provide more detailed explanations
and examples.

\begin{table*}[h!]
  \begin{center}
    \caption{Most-notable technology examples with anti-satellite capability}
    \label{tab:classic-threats-examples}
    \begin{tabular}{ p{0.2\textwidth} | p{0.75\textwidth} }
      \toprule
      Technology & Examples \\
      \midrule
      \multirow{3}{*}{direct ascent vehicle} & \textit{Prithvi Delivery Vehicle Mark-II}, tested successfully in 2019,creating limited amount of debris~\cite{Tellis19}\\
      & \textit{Operation Burned Frost}, removal of defunct satellite conducted by SM-3 vehicle~\cite{Blount09}\\
      & Taking down of~\textit{Fēngyún 1C} defunct satellite by \textit{SC-19} intercept vehicle~\cite{Mineiro08}\\
      \midrule
      \multirow{6}{*}{co-orbital vehicle} & \textit{X-37B} USAF unmanned spaceplane~\cite{X37B}\\
      & \textit{Chongfu Shiyong Shiyan Hangtian Qi} reusable test spacecraft~\cite{ChinaSpaceplane}\\
      & \textit{Aolong-1} active debris remove vehicle~\cite{Aolong1}\\
      & \textit{Kosmos 2519, 2521, 2523} small interceptor satellites performing randez-vous and proximity demonstrations~\cite{Kosmos2519}\\
      & \textit{Kosmos 2543} deployed \textit{Kosmos 2542}, which in turn started to spy on U.S National Reconnaissance Office's US245 (with the NRO satellite performing maneuvers to steer off the adversarial inspector)~\cite{Kosmos2543}\\
      & \textit{SJ-17} experimental satellite performed very close proximity operations on dead and operating Chinese geostationary satellites, showing capabilities for both peaceful and hostile actions~\cite{SJ17}\\
      \midrule
      projectile deployment & \textit{BX-1} deployment from \textit{Shehzhou-7} mothership and \textit{ISS} pass within 25~km (seemingly out of control)~\cite{BX1}\\
      \midrule
      high altitude nuclear detonation & \textit{Starfish Prime} was a nuclear explosion conducted at altitude 400 km, creating both electromagnetic pulse and artificial radiation belts. At least 6 satellite failures, that occurred in months that followed, are attributed to this experiment~\cite{Starfish}\\
      \midrule
      high energy directed weapons & operational laser systems like \textit{Peresvet}~\cite{Peresvet} or \textit{Sokol Eshelon}~\cite{Sokol}. The latter is an origin of successful (but temporary) blinding of Japanese Earth observing satellite AJISAI~\cite{Sokol2}\\
      \midrule
      communication eavesdropping & \textit{Olymp-K / Luch} has been
      launched in 2014, as communication satellite. However, instead of maintaining
      its position and performing typical operations, it started to wander around GEO
      belt. As of 2020 it has shifted its position about 19 times \cite{Harrison21},
      targeting mainly communication satellites, both military and commercially
      operated. Both, traversing of the GEO belt and close proximity operations
      sparked accusations of espionage and hostility (communication eavesdropping,
      inspection) \cite{Gruss15, Leicaster19, Tran18}.\\
      
      \bottomrule
    \end{tabular}
  \end{center}
\end{table*}

It is understandable that most of the community attention will be drawn to
spectacular weapons, capable of tracking and physically destroying or, at least,
damaging space systems. As mentioned before, there is not much that can be done
about preventing or avoiding such attempts or outmaneuvering them.
Yet, thankfully, we don't see many of such events, which as of to date 
are limited to technology demonstrations and targeted only to a nation's own vehicles.

There are several reasons for that: First, successfully intercepting
or damaging a rival's orbital vehicle is considered an attack to that
rival's territory. Space faring nations have established a balance of
powers agreement, threatening retaliation, including between the two
large defensive packs, in case space vehicles are attacked. The
aggressor must therefore expect retaliation on earth, which might well
ignite a spiral of uncontrollable escalation between the parties.

Second, physically destroying a vehicle multiplies the debris that must be
tracked and avoided because it would otherwise destroy further space adding
even more debris. The ultima-ratio of this effect is, introduced in~\ref
{ssec:sustainability}, Kessler's syndrome and will render space unusable for
generations to come, cutting technologically advanced space-faring nations from
the military and civilian applications that space provides.

Third, but of minor importance due to weak enforcement mechanisms,
international space treaties would be violated that aim at regulating
a safe, secure and sustainable use of this domain.

In developed countries, there is no real interest in actually risking
the destruction of one's own infrastructure by clouds of debris or
even preventing future use of such orbits for decades or until
clean-up technologies become available. The kinetic hit to kill philosophy in
space is clearly a double edged sword. In fact, weapons enabling
near-space existential threats work in a way like nuclear weapons ---
they are capable of affecting everyone including those who yield
them. In a similar fashion, like nuclear warfare, Mutually Assured
Destruction worked so far to stabilize and build the balance between
powers (however not making the world a better place). The risk of
rendering near-space unusable has a chance to push the space
powers rivalry into domains of non-existential threats.

The peaceful utilization of outer space is also a concern of the
United Nations (UN), creating the UN Institute for Disarmament
Research (UNIDIR) for monitoring and classifying all aspects related
to international security and to assist in disarmament
processes. UNIDIR hast spent more than 40 years evaluating technology
development and tracking global activity in the context of space
militarization. The Prevention of an Arms Race in Outer Space (PAROS)
program was very successful in increasing and widely spreading
awareness of threats and risks associated with placing weapons in
orbit, much less in preventing space powers in exercising their space
access and utilizing their denial capabilities \cite{UNIDIR5}.

However, space powers and, as we shall see, also non-space powers, less
developed countries or even terrorist groups do not need to target space
systems in a direct kinetic manner. Cyber- and electronic warfare allow for
significantly more subtle, but still highly effective attacks that are much
harder to attribute~\cite{UNIDIR3} and that require significantly less advanced
equipment. If the the careful reader reviews the threat summaries referenced in
this chapter, she will realize that majority of those surveys do not capture
the vast possibilities of cyber or electronic attack on space infrastructure,
which contributes to serious underestimation of, probability and the
criticality of such event.
Hacking a satellite grants adversaries control over the payload system
and the applications it supports. Through it, they may spoof GNSS signals,
disrupt, redirect, or manipulate communication, or use on-board
systems for espionage. Hacking into the platform system or, for that
matter, into a satellite's controlling ground station, grants full
control over the satellite, including its orbit and mass. The latter
two turn satellites into projectiles and kinetic weapons. The
equipment required to reach satellites in orbit is relatively simple
and, in particular, does not require capabilities of deploying or
maintaining vehicles in space. An immediate conclusion from the last
two aspects is obviously that threats of causing physical damage and
ultimately Kessler's syndrome are not limited to developed countries,
but also to countries and organizations that may not be as dependent
on space or that may even benefit from the inaccessibility of this
domain. In particular this includes asymmetric threats, which
are naturally immune to the threat of Mutually Assured Destruction.
%
Very few trusted and, proven to be effective, countermeasures exist against
cyber and electronic attacks, especially when compared to kinetic domain
(see Table~2 in \cite{Harrison21b}). It is not that the adequate countermeasure
cannot be conceived. The problem lies in, perception of the threats, which we
aim to change. 




\section{Cyber-physical threats to space systems}
\label{sec:space-cps-threats}
\st{general take on what threats are?}
\st{who could interfere with space systems and why?}
Spacecraft, space probes and space vehicles are cyber-physical systems:
Satellites operate remotely, either alone or, to an increasing extent, in
networked groups, but always in contact with ground stations, leveraging
algorithms on their computer systems to control actuators, propulsion and
sensors in order to fulfill their missions.
As such they can fall prey to all threats more traditional
cyber-physical systems (CPS) are exposed to, in addition to those
originating from the harsh environment in which they operate and those
relayed from the ground stations they communicate with.  Concerns
include classical security and dependability questions, like the
confidentiality and integrity of sampled or relayed data, or
the availability of subsystems, but also more specific ones, like misuse
of the vehicle, including as a projectile or impactor \cite{CyberThreat}.

In the following, we analyze the threat vectors spacecraft and space
infrastructure are exposed to, the agents that exploit them and give
examples of threats at various stages of the mission.
Fig.~\ref{fig:space-systems-nodes}) gives an overview of the units
involved in deploying and operating spacecraft, which may be
compromised by adversaries and exploited for attacking other units
\cite{VulnThreatVProb}.



\subsection{Terms and definitions}
\label{ssec:space-threat-vectors}
\emph{Threat vectors} describe possibilities through which agents may gain
access to system assets or resources \cite{VulnThreatVProb}.  A
threat, as defined in the seminal work of Singer~\cite{Singer58}, is a
product of the estimated capabilities of malicious actors and their
intent.
To defend a spacecraft against them, as many threats as possible need
to be identified and evaluated against the abilities of potential
malicious actors. Known weaknesses must be fixed, but since both
identification and patching remain incomplete processes, spacecraft
and supporting infrastructure should ideally also be prepared to
tolerate threats, such as partially successful attacks, and return to a
state at least as secure as initially. In other words, they must be
resilient to accidental and intentionally malicious faults.

A \emph{vulnerability} is a weakness or flaw in the system, its
configuration or operation, that might be exploited when reachable by
adversaries to gain an advantage over the systems and ultimately
compromise it~\cite{VulnThreatVProb}.

Exploited vulnerabilities lead to \emph{faults} in the system, which
may manifest in errors and ultimately lead to failure of the
system~\cite{Avizienis04}. We distinguish accidental from
intentionally malicious faults. Whereas the natural processes that
cause the former, the probability distributions they follow and their
low-level effects are well understood, this is not the case for the
latter. Whether, when and how frequent intentionally malicious attacks
can be mounted are often just a question of the power an adversary can
muster and of the incentives it has. Some adversaries operate with the
resources of nation states, directly provided to them.



The situation is further complicated by a vast variety of weaponized
software being available for analysis, re-purposing and deploying on
the adversary-owned systems, but more importantly on systems of
innocent users that have been compromised.
%
%

The proliferation of commoditized, low-cost space platforms, with
almost no trade restrictions, common architectures and COTS
technologies foreseen for NewSpace amplify this situation and expands
the window of vulnerability for attacks~\cite{Wrightson14}.

\subsection{Agents}
\label{ssec:threat-agents}
The agents opposing space systems include governmental, military or
commercial actors, but also individuals or organizations that
undertake attempts to explore selected threat vectors.  Since space
systems require multidisciplinary knowledge and the equipment and
software comprising such systems used to be rather rare and not easily
obtainable, it was possible to exclude the occasional hacker from this
list of adversaries. Other than that, the full spectrum of adversaries
has to be expected \cite{Insider, CCSDS-350-1-G3}:
\begin{itemize}
  \item sophisticated individuals
  \item insiders / untrustworthy or careless personnel
  \item competitors / dishonest or careless business partners
  \item hacktivists
  \item criminal organizations / guerrillas
  \item nation states
  \begin{itemize}
    \item state backed organizations
    \item intelligence
    \item military
  \end{itemize}
\end{itemize}


Before analyzing the threats in further detail to which space systems are
exposed, let us review the evidence we have from existing attacks, evaluated in
section~\ref{sec:attacks}.





\section{Review of known attacks}
\label{sec:attacks}
\st{
Space system are inherently vulnerable due to their nature. Some worry about
military counterspace ops, that can cripple whole nations, but same effects
might be created by persistent hackers, and the spectrum of possible bad things
that might happen to space system in cyber-physical domain is huge.
Following attacks review shall serve as the proof we're vulnerable.
}

\begin{table*}[h!]
  \begin{center}
    \caption{Summary of publicly known space infrastructure security incidents \cite{Fritz2013, SSI19, Horowitz15}}
    \label{tab:incidents-summary}
    \begin{tabular}{ p{0.05\textwidth} | p{0.25\textwidth} | p{0.6\textwidth} }
      \toprule
      Year & Incident & Remarks \\
      \midrule
      1998 & ROSAT & Scientific satellite payload permanent failure coincidental with cyber-intrusion to mission control center, incident report classified \cite{Epstein08, Talleur99} \\ 
      1999 & Skynet & British military communication satellite allegedly taken over and ransom requested, lack of solid, public, evidence of incident \cite{Skynet99a, Skynet99c}\\
      2000 & GPS jamming during military trials & British and US tank had navigation problems during Greek trials. GPS jammers deployed by French security \cite{Grau01}\\
      2003 & Ames Research supercomputer shut down to halt intrusion & Swedish national persecuted, estimated costs > 1MUSD \cite{NASA-OIG-intrusions} \\
      2003 & TELSTAR-12 uplink jamming & TELSTAR-12 uplink was jammed, by source located in Cuba, during Operation Iraqi Freedom to prevent Voice of America broadcast over Iran~\cite{Corneau03} \\
      2005 & Sri Lankan rebels hijack satellite communications & Liberation Tigers of Tamil Eelam broadcast pirated TV \& radio services to several countries \cite{TamilTigers07} \\
      2006 & Data breach and multiple intrusions & NASA forced to block emails, Shuttle operations plans leaked \cite{Boyle06, Hack06} \\
      2007 & Landsat-7 & First unauthorized attempt to access the space segment \cite{USCC11, USCC15} \\
      2008 & Landsat-7 \& Terrasat EOS interference & Very well documented hack attempt, large sophistication of adversary \cite{USCC11, USCC15} \\
      2008 & Worm infecting laptops on ISS & Brought by a Russian astronaut on Windows XP laptop. Malware quickly spread among other computers (although mission-critical equipment was safe) \cite{ISSvirus08a, ISSvirus08b} \\
      2009 & JPL data breach and malware spreading in NASA mission networks & Theft of 22GB of export-restricted data; thousands of connection set to external networks \cite{NASA-OIG-intrusions-b}\\
      2009 & BBC broadcast in Farsi disrupted & Telecommunication satellite jammed \cite{Horrocks09}\\
      2009 & NASA Goddard Center information leaked & Paid Earth imagery datasets posted online for free \cite{NASA-OIG-intrusions}\\
      2010 & GPS jamming by N. Korea & Multiple locations affected in S. Korea including Incheon International Airport. Aircraft had to rely on alternative navigation instruments. Incidents repeated couple of times in following years \cite{Mizokami16}\\
      2010 & NASA intrusions & Data destroyed or access restricted, 0.5 MUSD damage to Atmospheric Infrared Sounder (AIRS) program \cite{JPL12} \\
      2011 & NASA JPL breach & Hackers gained full access to JPL systems \cite{Finkle12, JPL12} \\
      2011 & European communication satellite jamming & Deutsche Welle jammed on DeHotbird 8 satellite \cite{Goertz11} \\
      2011 & NASA ISS command and control data leak & An un-encrypted NASA laptop was stolen. It contained the command sets, as well as, control algorithms for ISS \cite{Samples11, JPL12} \\
      2011 & JAXA H-2A Transfer Vehicle design leak & Virus infected laptop containing critical data \cite{JAXA11} \\
      2012 & NASA and ESA identity and authentication data hacked and published & Around one thousand employees personal information leaked and posted in internet \cite{Protalinski12} \\
      2012 & JAXA Epsilon rocket design leak & Virus infected laptop containing critical data \cite{JAXA12} \\
      2014 & DLR breach and data theft & Targeted malware found across DLR computers. Theft linked to China APT groups \cite{DLR14} \\
      2014 & Mulitple channels broadcast disrupted over Ethiopia & Arabsat telecommunication satellite jammed \cite{deSelding14} \\
      2014 & NOAA satellite weather imagery service disrupted & Data flow from satellites affected by hack attributed to Chinese APT, systems forced offline \cite{NOAA16, Gruss14} \\
      2015 & \textit{Turla} satellite communication links hijacks & \textit{Turla} hacker group with links to FSB - hijacking internet services of older commercial satellites \cite{Zetter15, Lennon15, Tanase15} \\
      2015 & \textit{APT28} hacked French TV5Monde television & A professional, coordinated attack that disabled the TV broadcaster for couple of hours. It took months to fully replace destroyed equipment and return to regular operations \cite{Corera15}\\
      2018 & JPL intrusion & 500 MB of critical documents leaked, unauthorized access to deep space network, operations affected for many months, \cite{Winder19, JPL19} \\
      2018 & malware in ISRO launch segment & suspected, ISRO named it false positive \cite{Mithun18, XtremeRAT, ISRO18} \\
      2018 & DoD contractors hacked & Security breach with the possibility to exercise the control over satellite by hackers, data traffic disruptions. Additionally confidential design data on submarines and high fidelity satellite imagery stolen \cite{Starks18, CNBC18}\\
      2019 & Advanced GPS signal spoofing in China & Ships GPS positions, reported by maritime satellite AIS system \cite{Harris19, Harrison21}\\
      2019 & Successful attack on autonomous car navigation by GPS spoofing & \cite{Regulus19}\\
      2020 & Worldwide advanced GPS signal spoofing & Ships located physically in waters near Norway, Libya, Malaysia, and Russia reported via AIS to sailing in circles off the San Francisco coast\cite{Spocchia20, Osborne20}\\
      \bottomrule
    \end{tabular}
  \end{center}
\end{table*}

\subsection{ROSAT failure}
\label{ssec:rosat-failure}
The earliest mention of counter satellite activity can be traced back
to 1998 when ROSAT failed. ROSAT was an American-English-German
scientific satellite that first experienced a malfunctioning reaction
wheel used for attitude determination and, as a consequence, turned
its instruments directly towards sun which destroyed
it~\cite{Epstein08}.
The ROSAT platform was plagued by faults and issues from its early
days on~\cite{Harland05}.
The possibility that hackers might be responsible was raised a decade
later by T. Talleur~\cite{Talleur99} in a confidential report along
with a report on other malicious activities in the NASA
networks. While, at the present moment, the original article is no
longer available on-line (including in the Internet Archive), and the report
is not publicly available for obvious reasons, the event has been
reported by respected security researchers~\cite{Schneier08} and US
security NGOs~\cite{Rosenzweig12}. Currently the article is backed up
on the author's blog \cite{Epstein08b}.

Since there is no direct evidence for the malicious activity to be the
root cause of final failure of ROSAT and not one of the existing other
other plausible root causes, this incident has to be taken with a
grain of salt.
\rg{anything to add on-like stolen data, ROSAT was platform similar to
  others?}

\subsection{Skynet}
\label{ssec:skynet-hack}

As reported by Reuters and Time \cite{Skynet99a}, in February 1999,
one of United Kingdom's Ministry of Defence military telecommunication
satellites unexpectedly changed its orientation. Soon after, as the
initial story goes, the MoD received a ransom request to gain back
control over a critical piece of infrastructure. Early March that
year, all allegations of a satellite hack and subsequent service
interruption were denied by officials \cite{Skynet99c} and, in the
end, ridiculed \cite{Skynet99d}. Whatever really happened then, the
event sparked panic among military and intelligence personnel, to the
extent that some (mis)information leaked to the public. If such an
event actually took place, it would be evidence of the significant
technical sophistication of an adversary or of a significant security
misconduct on the defending side \cite{Skynet99b}. It is not very
probable, but in the light of other, better documented cases, also
potentially feasible.

\subsection{Landsat and Terrasat EOS}
\label{ssec:terrasat-hack}
Turning to better documented incidents, 
in late 2007 and 2008 two US government remote sensing, Earth
observation satellites became subject to adversarial activity of
unattributed origin (with presumptive evidence leading to a global
competitor). In October 2007, Landsat-7 experienced about 12 minutes
of interference, which was only discovered following the analysis of a
subsequent event 9 months later. This second attempt to take over
control over satellite was also not successful. In June 2008, Terra
EOS AM–1, became subject to about 2 minutes of interference. The
adversary, managed to complete all steps required for obtaining
command authority over the satellite but refrained from issuing rogue
commands. Four months later, the same satellite experienced a 9 minute
hostile take over attempt, with all the steps required to take over
control completed and again attackers restraining themselves from
issuing rogue commands. Those events, clearly show the scale of threat
and sophistication, as well as, the fact that malicious actors possess
advanced knowledge on system operation details. Unlike the previous
examples, the above events are credibly documented by Economic and
Security Review Commission reporting to U.S. Congress \cite{USCC11,
  USCC15}.

\subsection{NASA JPL breaches}
\label{ssec:jpl-hacks}
In November 2011, NASA's Jet Propulsion Lab discovered compromised
accounts of several highly privileged users. The hackers had full
systems access, enabling them to copy, modify and delete files as well
as to create new user accounts. They could as well have uploaded
malware for further exploitation of the NASA networks
%
%
\cite{Finkle12}. In the course of the investigation it was revealed
that for the past 2 years, NASA was subject to significant adversarial
activity, aiming at accessing the internal data networks, causing, both
IP leaks and interruption of operations \cite{JPL12}.  Unfortunately,
the recommended strengthening of NASA JPL's defense posture was
insufficient, as information that surfaced in June 2019 indicated
another heavy network security breach, that happened in the Agency's
flagship laboratory, 14 months earlier. This time an unauthorized
Raspberry Pi microcomputer was found plugged into the facility
network, providing cyber-access for adversarial activity, including 
access to confidential documentation and to the Deep Space Network (DSN) - an
array of radio-telescopes used for ranging, telemetry acquisition and
remote control of exploration probes traversing the Solar System. The
scale of that compromise shall be alarming, as officials admitted, the
hackers gained access to the gateway enabling them to take over or at
least affect the mission control centers, including to those related to the human
spaceflight program \cite{Winder19}.  Malicious activity went
undetected for 10 months. Initial damage assessment mentioned a leak of
500 MB of highly sensitive information, some under ITAR restrictions,
and vast, long lasting disruption of network operation, including DSN
and connections to other NASA sites \cite{JPL19}.

\subsection{NOAA}
\label{ssec:noaa-hacks}
In October 2014, the flow of meteorological data from satellites
operated by U.S.\ National Oceanic and Atmospheric Administration has
been temporarily disrupted by an internet-sourced attack. Some
forecasting services were disrupted, while the systems were under an
unscheduled maintenance \cite{NOAA16}. The incident analysis report
stated that the organization did not sufficiently implement security
requirements despite several audits that expressed this
obligation. This lack of protection, left critical communication
systems vulnerable, namely the satellite data feed and interfaces to
other parts of critical infrastructure, including military
\cite{Gruss14}.

\subsection{Indian launch site malware infection}
\label{ssec:indian-launch-malware}

In December 2017, a malware \textit{XtremeRAT} has been found by
Indian and French independent researchers in India's Space Research
Organization (ISRO) Telemetry, Tracking and Command Networks (ISTRAC)
used for control and support of launch activities from the launcher
ignition up to payload orbital injection.  The malware could even have
been present on a computer that was directly involved in launch
operations \cite{Mithun18}. \textit{XtremeRAT} is a standard and
widely available offensive tool used for accessing and taking control
over a victim's system, often targeting critical infrastructure
\cite{XtremeRAT}. ISRO, after conducting an internal investigation, declared the
 incident as a false positive. The organization pointed to the fact that ISTRAC
 mission critical systems are air-gapped and thus secure against this type of
 adversarial activity \cite{ISRO18}. Air gapping disconnects a computer from
 all external networks. History has shown that this approach is a challenge for
 hackers, but the one that can be, eventually, and, spectacularly, beaten as in
 famous \textit{Stuxnet} malware~\cite{Albright10}. However, having to maintain
 wireless connection to space vehicles, air-gapping space systems naturally
 remains incomplete, as the, reverse, space-to-ground infection path
 feasibility shall be also considered.

\subsection{GPS spoofing}
\label{ssec:gps-spoofing}

There is a significant number of cases where GPS signals have been
either jammed or spoofed, a fact which is not very surprising given
how weak those signals are when received on the ground, easing malicious
interference. It is also not very surprising that adversaries have an
incentive to mount such attacks, given how large the dependence of our
society, economy and military on satellite navigation systems is.  The
GNSS receiver market is more than 97 BUSD worldwide, which is about
one quarter of the global space economy in 2019 \cite{Bryce20a}.
GNSS signals have been subject of extensive testing against jamming,
meaconing or spoofing techniques for couple of years
already~\cite{Buesnel16}), however, today's receivers are still
extremely easy to be tricked into false position, velocity and time
reporting. \textit{GPS crop circles phenomena}, sign of advanced GPS
spoofing technology, has been first recorded in 2019 in several spots
in China. As data received through the satellite Automated
Identification System (AIS,\cite{AIS}) indicate, ships that entered
the spoofing area reported their positions as sailing in circles
around arbitrary locations on land \cite{Harris19, Harrison21}. In the
following year, similar incidents happened on a worldwide scale, where
a couple of ships around the world have been tricked into recognizing
(and further, reporting through their AIS) its position off the coast
of San Francisco \cite{Spocchia20, Osborne20}. It is worth mentioning,
that GPS spoofing is not only limited to close proximity of
interfering equipment. Using relatively cheap and widespread
technology (e.g., Software Defined Radio and open source software
stacks \cite{Myrick17, GPSspoof-codebase}) spoofing ranges of up to
tens of kilometers can be achieved \cite{Hambling17}. Using a large
number of jamming devices, areas of the size of entire countries can
be equipped with counter-GNSS technology \cite{Pole21}. In the near
future, with more advanced technical deployments, like
\textit{Ekipazh} nuclear powered in-orbit electronic warfare
satellites \cite{Hendrickx19, Harrison21}, GNSS signal jamming or
spoofing (or communication links disruption) could possibly affect
much larger areas.

\subsection{Communication links hacks: Turla and APT28}
\label{ssec:comm-link-hacks}
Since a significant part of existing space telecommunication
infrastructure is based on analogue relays (\textit{bent-pipe}
concept) it is easy to jam inputs of such satellites. Essentially, all
the adversary needs is to know the uplink frequency (which is public
information) and a strong, directional source of interference.
Unless, more complex, digital, regenerative payloads, are used, the
only countermeasure to this attack is to track down the sources of
interference and shut them down physically. There are cases were
satellite communication infrastructure has been either target or
conduit of more sophisticated attacks. In 2015 a \textit{Turla} group
has started a large-scale hijacking of satellite internet links in
order to disguise their activity and hide their physical presence
(since end-user terminals can be placed anywhere within the satellite
service beam). Some of this activity was simply purchasing
bidirectional links as any customer would do. However, to avoid the
significant cost involved with this, \textit{Turla} started to spoof
packets of DVB-S based Internet system users. It was easy, because
unlike its successor DVB-S2, DVB-S is not encrypted \cite{Zetter15,
  Lennon15, Tanase15}.  The same year, TV5Monde experienced a
devastating cyberattack on a few of its broadcast facilities by
hacker group \textit{APT28}. The service (all 12 channels) was down for
a couple of hours. It may not sound much, but broadcasters have contract
obligations on signal availability and contract cancellation could
jeopardize the company's existence. The attack was sophisticated and well
prepared. Months earlier adversaries mapped TV5Monde networks,
understanding how the broadcast process works and what equipment is
involved. For grand finale, a malicious software was deployed,
targeting critical broadcast devices and causing permanent hardware
damage \cite{Corera15}.


\section{Threat Vectors}
\begin{figure*}
  \includegraphics[width=\textwidth]{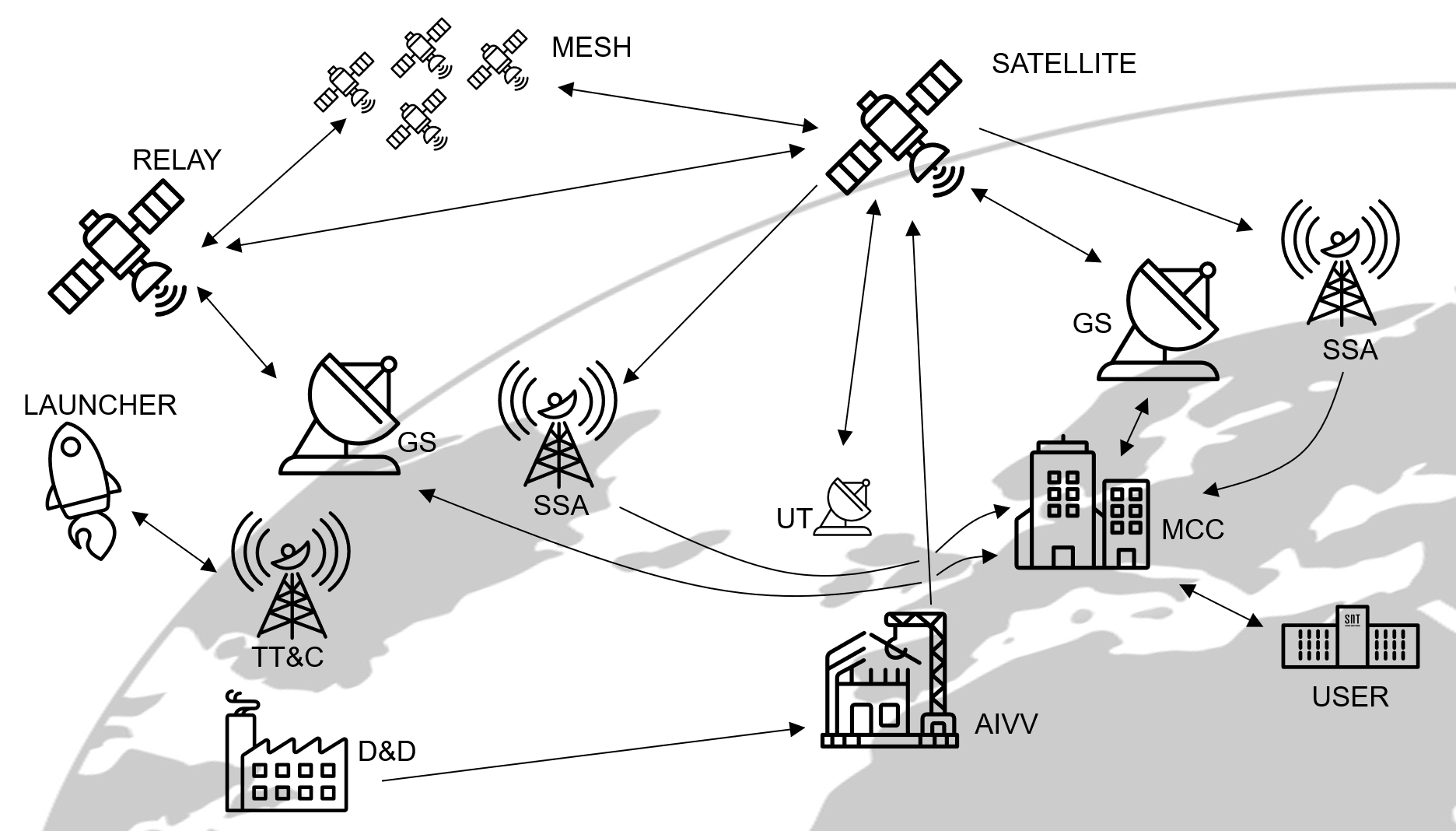}
  \caption{Overview of space systems nodes}
  \label{fig:space-systems-nodes}
\end{figure*}

\st{
After showing the greater context and the dangers,and going through threats in
general, here, I, narrow the down to cyber-physical classified wrt confidentiality,
integrity, availability for all the segments and aspects of the space system
existence.
}
In the following subsections, we now analyze systematically the
threats to which space systems (see Fig~\ref{fig:space-systems-nodes})
are exposed to in the individual phases of their lifecycle. We
distinguish physical from cyber attacks, but highlight also where
cyber attacks influence the physical world. We investigate to which
extent a successful manifestation of these threats affect
confidentiality or integrity of data, or, the availability of the spacecraft itself, before we provide in
Section~\ref{ssec:space-threats-scenario} a general scheme for the cyber-physical
attacks on the space systems.

\subsection{Early phase attacks}
\label{ssec:early-phase-attacks}

\begin{table*}[h!]
  \begin{center}
    \caption{Classification of threats to space systems - supply chain, assembly, integration and tests~\cite{CCSDS-350-1-G3, Bailey19, Bailey20, Oakley20, Harrison21}}
    \label{tab:threats-classification-1}
    \begin{tabular}{c l c | c | c }
      \toprule
      \multirow{2}{*}{} & \multirow{2}{*}{} & \multicolumn{3}{c}{Affecting} \\
      \cmidrule{3-5}
      & & supply chain & assembly \& integration & tests \\
      \midrule
      \multirow{6}{*}{\rotatebox[origin=c]{90}{loss of}} & \multirow{2}{*}{confidentiality} & design theft & documentation theft & test plan theft\\
      & & specification theft &  & test results disclosure\\
      \cmidrule{2-5}
      & \multirow{2}{*}{integrity} & component tainting & documentation modification & test specification modification \\
      & & design modification &  & test equipment settings modification \\
      \cmidrule{2-5}
      & \multirow{2}{*}{availability} & component supply disruption & documentation deletion & test equipment unavailability \\
      & & design deletion  & integration facility unavailability & test results deletion \\
      \bottomrule
    \end{tabular}
  \end{center}
\end{table*}

Before space infrastructure is even built and deployed in space, it
undergoes long process of design, modification, assembly, integration
and extensive testing. This is almost always a collaborative effort,
involving many parties bringing different expertise, know-how and
facilities. Hence, a large attack surface must be expected. It is also
very important that as a single security compromise at those early
stages, can not only stop or delay the program, but it may embed
vulnerabilities that lay the foundation for later, more elaborate
attacks on the developed ground or space system.

Attack vectors in this early phase affect satellite system design and
development houses, component supply chains --- D\&D in
Fig~\ref{fig:space-systems-nodes} ---, assembly (manufacturing of
mechanisms, electronic circuits), integration (assembling whole
platform, payload, satellites), verification and validation
(conducting the functional and environmental tests on integrated
equipment) --- AIVV in
Fig~\ref{fig:space-systems-nodes}. Vulnerabilities embedded at this
stage may remain in existence well beyond the time when the satellite is finished, deployed into orbit and operated there.

At the early stage of system development, supply chain attacks offer a
first interesting opportunity for adversaries to learn about the
systems' capabilities. Moreover, state-backed agents may inject
trojan-horses and other malware, both, at software and at hardware
level, which constitutes an invaluable opportunity for further
compromise of units after they become operational.

If the supply chain is of military grade, such attacks are not trivial, but
still possible~\cite{Skorobogatov12}. If the supply chain includes open source
designs and components, as it is increasingly popular New Space (\cite{CFS_a, CFS_b, FreeRTOS, LibreCUBE, LibreSpaceFoundation}), \ tainting of the
design or its components seem easy at first glance, but are less trivial if
they should both evade public analysis and the later validation process.

Similar activities, dealing mainly with documentation theft,
unauthorized design alterations or simply deletion have to be expected
at the integration stage.
Some of the documents that are typically prepared in the course of
developing satellite systems are more crucial than others. For
example, Failure Modes, Effects and Criticality Analysis (FMECA)
describe known failure modes and their assessment. It explains what is
to be expected to go wrong and how the system is prepared to face the
challenge.
Such a document falling into the hands of adversaries informs them
about the weakest link, which may guide them in the selection of the
subsystem they target.
More interesting even are the failure modes and effects which have evaded
FMECA, as they pave the way for exploiting undocumented
(and therefore likely untested) behaviors~\cite{Oakley20}.

The test phase is another, very attractive epoch in the development life cycle to
learn about or affect space systems. Knowledge about test procedures and
methods for satellite system verification and validation, tells how a satellite
system shall be utilized and which behavior is to be expected and which
behavior constitutes anomalies. Adversaries could attempt to avoid exhibiting
the latter to disguise their activity. Moreover, the test procedures themselves
can be used to attack the space system, for instance by manipulating test
equipment to leave critical functionality untested or by removing important
steps from the procedure itself.
%
%
It is also worth noting that failure of test equipment can destroy systems or
delay their deployment. Even if the system under test is not destroyed, but
pushed beyond the agreed range of operation, common procedure is that they have
to be considered as broken and should not be deployed in space.
%
%
The above threat vectors are summed up in
Table~\ref{tab:threats-classification-1}.

\subsection{Ground segment attacks}
\label{ssec:ground-segment-attacks}

\begin{table*}[h!]
  \begin{center}
    \caption{Classification of threats to space systems - ground segments~\cite{CCSDS-350-1-G3, Bailey19, Bailey20, Oakley20, Harrison21}}
    \label{tab:threats-classification-2}
    \begin{tabular}{c l c | c | c | c }
      \toprule
      \multirow{2}{*}{} & \multirow{2}{*}{} & \multicolumn{4}{c}{Affecting} \\
      \cmidrule{3-6}
      & & launch & ground station & mission control & space traffic management\\
      \midrule
      \multirow{6}{*}{\rotatebox[origin=c]{90}{loss of}} & \multirow{2}{*}{confidentiality} & payload disclosure & eavesdropping & eavesdropping & tracking \\
      & & tracking & tracking & (loss of encryption) & \\
      \cmidrule{2-6}
      & \multirow{2}{*}{integrity} & trajectory modification & masquerading & modification of data & man-in-the-middle attack \\
      & & GNSS meaconing / spoofing & message replay & modification of commands & tracked objects catalogue modification \\
      \cmidrule{2-6}
      & \multirow{2}{*}{availability} & loss of launcher & denial of service & denial of service & deletion of tracked objects catalogue \\
      & & no separation & jamming & facility unavailability  & observing station disablement \\
      \bottomrule
    \end{tabular}
  \end{center}
\end{table*}

The ground segment forms a large part of space infrastructure. Ground
stations (GS on Fig~\ref{fig:space-systems-nodes}) are responsible for
communicating with space systems. Mission control centers (MCC in
Fig~\ref{fig:space-systems-nodes}) initiate and supervise the systems'
operations and disseminate gathered data or provide other services to
users (e.g., distribution of intelligence data, earth imaging or
feeding in the broadcast signals from stations).
The exact architecture of this segment will vary between systems but
also depending on type of activity performed by the space systems.  An
important part of the ground segment is also the space traffic
management or space situational awareness (SAA on
Fig~\ref{fig:space-systems-nodes}) infrastructure, which is comprised
of ground facilities and responsible for tracking both, active and
passive, space systems. Data processing centers archive and distribute
ephemerids and conjecture assessment alerts to satellite system
operators to warn about the risk of collision.

{\bf Ground stations} are often considered choke points of the space
system infrastructure~\cite{CCSDS-350-1-G3}, and are therefore most
obvious and desirable targets of cyberattacks. 
%
%
%
%
Eavesdropping, at first, is useful for identifying the satellites a
given facility communicates with. For ground stations this correlation
is straight forward by means of which satellites the station can
see. This is of course provided data is not relayed using satellites
in this window to reach to other satellites.

For {\bf mission control centers}, if linked with a distributed network of
ground stations, more elaborate analysis is required. One needs to
find out at least which ground stations are in the network. Breaking
confidentiality at this stage helps adversaries find 
out about physical (at GS i.e. RF signal encoding, framing) or higher
lever (at MCC) communication protocols.
Better prepared attackers can attempt to impersonate legitimate ground
stations to communicate on their own or at least replay recorded
communication. If attackers are able to masquerade as a mission
control center (e.g., by stealing the center's authorization keys),
they may modify or forge commands to obtain unconstrained access over
a satellite system.  As of today, satellites have no means to verify
the sanity and safety consequences of received command
streams.
Hardest to mount would be attacks on GS or MCC availability, leading
to the unavailability of the targeted facility.
The spectrum of such attacks is very wide and spans from physical
attacks to cyberattacks tempering with equipment in ground stations
(antenna rotors, radome and rotor heaters, power outage). Cyberattacks
on GS and MSS may deny the service of such stations and classical
electronic warfare like RF jamming, saturating the station's receivers
with locally produced emission, prevent reception or tracking of the
satellite. Using directional RF beams, adversaries may even interfere
with commands legitimately sent by a ground station.

Other, often overlooked, components of the ground segment that are
however equally prone to attacks are the {\bf space situational awareness}
(SSA) facilities (laser and radar stations, telescopes) and the
digital infrastructure (i.e. data centers) that support them. SSA
serves (or will serve, as it is in an early stage of development) as
space traffic management backbone, granting active spacecrafts the
ability to avoid collisions with other, active or defunct spacecraft
and with debris. The criticality of such a service for the safety of
orbital operations cannot be underestimated as every failure adds to
the complexity of the task it fulfills.

Another, often overlooked threat vector are attacks on the {\bf
  post-processing infrastructure} of data relayed from satellites.
Mission Control Centers feed data from satellites to processing
facilities or to intermediary data operators, which process, store and
index data before disseminating the post-processed data to
end-users. A multitude of interfaces, network links, and facilities
are managed by third parties, offering a very large attack surface for
targeting the data stream.
Attacks include data theft, disclosure, modification, deletion or
corruption of data or metadata (communicated or stored), but also
denial of service attacks on the data storage
itself~\cite{Oakley20}.
%

Traffic information, satellite ephemerides or orbital conjunction
alerts, are distributed to satellite systems operators where this data
is used for orbital corrections and most importantly collision
avoidance maneuvers. Many of those information exchange systems are
still using solutions from the 1970's, like Two Line Element's sets
(TLEs~\cite{TLE}). If, as it is often the case, ephemeris data is
sourced from just one organization, this would give rise to 
man-in-the-middle type of attacks, fiddling with the TLEs
provided to satellite infrastructure operators, in order to
orchestrate unnecessary maneuvers, or worse, collisions.
Such an endeavor does not require significant processing power nor
sophisticated equipment~\cite{Pavur19}, which turns this attack into a
serious threat for space systems, in particular for terrorist
organizations and less developed countries, in particular as it
targets the limited resources satellites are deployed with. 

This threat has already been identified by international
organizations, such as CCSDS, addressing the extremely low security
concerns of TLE messages by designing a modern standard for orbital
information dissemination, better prepared to withstand
confidentiality and integrity attacks~\cite{CCSDS-orbital-msg,
  CCSDS-orbital-conjunction}). Similar effects can be achieved by,
less sophisticated, unauthorized deletion of selected ephemerides or of
critical collision alerts.

The threat vectors of this subsection are summarized in
Table~\ref{tab:threats-classification-2}.

\subsection{Launch segment attacks}
\label{ssec:launch-segment-attacks}

The launch segment comprises launch vehicles and tracking, telemetry
and command facilities which oversee the launch process. Since the
whole launch process is quite rapid, it will be quite difficult for
adversaries to launch elaborate attacks against the launched vehicle.

Attacks aim at revealing the launched payload, such as the type of
satellite and its capabilities, and the orbital injection parameters,
which can later be confirmed with the help of tracking capabilities in
the hands of the adversary.
Since the launch vehicle guidance is autonomous, malicious trajectory
manipulation would have to target on-board computers before launch,
which will be difficult, though not impossible.  Also the time-frame
of the launch process makes attempts to attack the launch trajectory
by spoofing the GNSS receivers rather unpractical, in particular
because many on-board guidance systems anyway rely on internal,
inertial measurement units.
Interesting attack targeting the launch process, would be to masquerade as a
legitimate ground facility in order to issue self-destruct or course change
commands, or to masquerade as a launch vehicle telemetry unit to report
anomalous readouts, tricking the launch safety officers into executing safety
procedures, which often lead to the destruction of the vehicle. Perhaps, the
simplest attacks on a launch systems are acts of sabotage leading to loss of
the launcher or preventing stages or payload separation (such events happen as
an accidental human error~\cite{Azriel13, Clark20}). In many cases, such
attacks require physical access to the vehicle, but some can also be mounted as
a result of a cyberattacks, manipulating on-board systems prior to the launch.

The described threat vectors are summarized in
Table~\ref{tab:threats-classification-2}.

\subsection{Space segment attacks}
\label{ssec:space-segment-attacks}

\begin{table*}[h!]
  \begin{center}
    \caption{Classification of threats to space systems - space and user segments~\cite{CCSDS-350-1-G3, Bailey19, Bailey20, Oakley20, Harrison21}}
    \label{tab:threats-classification-3}
    \begin{tabular}{c l c | c | c | c }
      \toprule
      \multirow{2}{*}{} & \multirow{2}{*}{} & \multicolumn{4}{c}{Affecting} \\
      \cmidrule{3-6}
      & & space: platform & space: payload & space: formations & user terminal \\
      \midrule
      \multirow{6}{*}{\rotatebox[origin=c]{90}{loss of}} & \multirow{2}{*}{confidentiality} & unauthorized access & unauthorized access & eavesdropping & eavesdropping \\
      &  & & & & tracking?? \\
      \cmidrule{2-6}
      & \multirow{2}{*}{integrity} & unauthorized access and commanding & unauthorized access and commanding & masquerading & data modification \\
      & & fault induction & fault induction &  & beaconing and spoofing\\
      \cmidrule{2-6}
      & \multirow{2}{*}{availability} & jamming & jamming & jamming & jamming \\
      & & platform failure & blinding & denial of service & service disruption \\
      \bottomrule
    \end{tabular}
  \end{center}
\end{table*}

As can be seen in Fig.~\ref{fig:space-systems-nodes}, the space segment is
comprised of satellites (which, as described in Sec.~\ref
{ssec:on-board-systems}, are traditionally divided into a platform and a
payload part), some of which assume the role of relaying units, and of groups
of interconnected satellites (constellations, formations, clusters). Probes and
deep-space exploration equipment would also fall into the satellite category.
The only difference is their the size, in particular of the antennas required
to remain in contact to ground and the amount of power devoted to transmission
for successfully communicating with earth.

Depending on the exact space system application the system's
architecture will differ significantly. The range of architectural
possibilities spans from one satellite, connected to one ground
station, to constellations of satellites, comprised of hundreds if not
thousands of nodes.  Constellation nodes increasingly become capable
of two-way communication with numerous ground-based user terminals,
but also, of communicating (node-hopping) with other
satellites within the constellation and of relaying information using
inter-orbit communication links. They are supported by an extensive ground
stations network, serving as an information exchange gateway but also
assisting in constellation control and management
purposes~\cite{Maral20}.

Breaking confidentiality in the course of an attack, regardless of
whether the attack pertains to platform or payload data, would be the
most common way of taking unauthorized advantage of a satellite. For
example, in the telecommunication domain, since satellites are used to
provide services to vast areas on Earth by means of relatively small
terminals, eavesdropping (or even integrity attacks by tampering with
messages) have a really low entry threshold~\cite{Maral20}. In fact,
many heavily proliferated systems, have none or very weak,
low-level encryption and authentication mechanisms, like DVB-S, with
significant, easily exploitable vulnerabilities~\cite{Pavur19b}. The
expectation is to provide security mechanisms at higher levels of
protocols~\cite{CCSDS-space-link-dvbs2}, but this requirement will
have to fight it's way through countless trade-offs as terminal power
consumption, processing capabilities and overall complexity will be
affected by encryption.

On the platform side, a similar process
takes place, but the vision of total loss or platform takeover
motivates better precautions, at least among the concerned, military and
governmental, as well as international organizations. Again, a lot of
civilian, especially scientific systems utilize CCSDS protocols,
ensuring standardization (and thereby availability of equipment) and
interoperability between different stakeholders and mission
participants.
An overview of how data systems (usually used in civilian exploration
probes or scientific or general purpose satellites) are organized can
be found here~\cite{CCSDS-space-architecture-from-data-systems}. For
example, the Space Packet Protocol
\cite{CCSDS-space-data-link-protocol} and the Space Data Link Protocol
\cite{CCSDS-space-link-protocol} do not contain any security measures
(such as authentication, confidentiality, or integrity ensuring
mechanisms) unless they are combined with the Space Data Link Security
Protocol \cite{CCSDS-space-link-security}, which is optional, although
encouraged. Obviously, in light of the examples shown in
section~\ref{sec:attacks}, one must also take into consideration that
encryption and authentication keys can be stolen or that a
vulnerability in the encryption algorithm, or its implementation, is
found, as has happened recently with SSL~\cite{Satapathy16}.
In such circumstances, adversaries not only become capable of
eavesdropping communication. They may also access the spacecraft's
internals and, if the adversary is sufficiently knowledgeable
(compare~\ref{ssec:early-phase-attacks}), issue unauthorized commands
to take over the vehicle (e.g., by installing new encryption keys) or
to inject faults. Such activities can result in satellite systems
switching to safe mode (reducing their functionality to what is
essential to survive, sometimes less) or loss of the system or part of
it (e.g., when adversaries points an Earth-observing telescope to the
Sun to permanently damage opto-electronic components). Either way,
availability of the system will be severely compromised.

Unfortunately, a direct access to on-board systems of a satellite, is not
required to severely compromise availability of space-based service, as
evidenced by numerous examples of electromagnetic interference attacks,
collected in Table~\ref{tab:incidents-summary}.

RF jamming for, i.e.  positioning system availability denial
requires unsophisticated transmitter, interfering at frequency of
interest with legitimate signals, saturating GNSS receiver inputs and
preventing correct reception of position. There are even works
available showing the possibility of selective GNSS signal denial,
providing similar results, but harder to
detect~\cite{Caparra18}.

Jamming the communication satellite uplink channels (used to receive the signal
to be further broadcast) is feasible, requires high power transmitters with
directional beams, but it also requires them to be positioned relatively close
to operator's transmitters (to target the high gain lobe of the receiving
antenna on the satellite). Jamming the downlink might be easier on technical
side (no pointing required, omnidirectional antennas and high power source of
interference signal is sufficient), but requires more jammers to cover area of
interest. Optical observation satellite payloads can be temporarily disabled by
using high energy lasers, in similar manner as radio links are jammed.

The future of near-space exploitation open up new ways of attacking the systems.
The LEO megaconstellations, which are being effectively deployed at the time of
writing this paper, assume that nodes in the network are heavily
interconnected.  Such set-up offers extra opportunities for malicious behavior
including eavesdropping by means of co-orbital vehicles (or masquerading as a
valid network node that can be used for message relay). More advanced attacker
could attempt to forge or to modify the messages in relay process, or simply
drop them at convenient moments~\cite{Oakley20}. While such possibility of
inter-satellite misbehavior can't be precluded, it requires a lot of technical
efforts and such high level of sophistication, so it is rather be implemented
for highest criticality assess (most likely in GEO, military or governmental
communication relays). On the other hand, despite still being on drawing
boards, satellite 5G communication networks are already scrutinized as opening
another venue of system abuse, through vastly proliferated IoT devices~\cite
{Gibson18}.

The vectors of attack presented above (and summed up in Table~\ref
{tab:threats-classification-3}) require elaborate knowledge about space
systems, their construction and operation. Then, if executed properly, attacks
of the above kind may allow adversaries to take over control, temporarily or
permanently compromise the availability of the spacecraft or even turn it into
an instrument of their purpose.


\section{Scenario for Cyber-Physical Threat to Space Systems}
\label{ssec:space-threats-scenario}

\begin{figure*}
  \includegraphics[width=\textwidth]{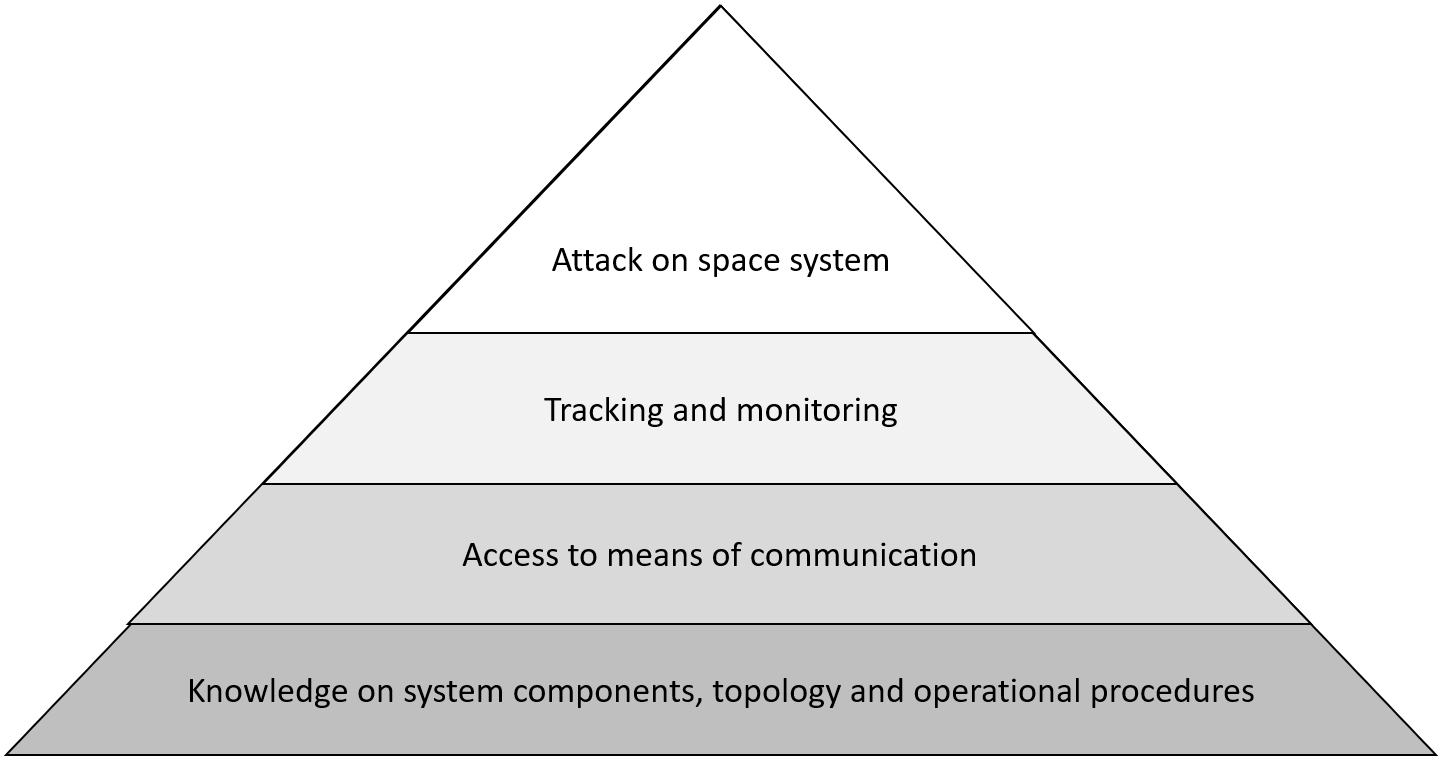}
  \caption{Stages of space cyber-physical system attack}
  \label{fig:space-cps-attack-pyramid}
\end{figure*}

Outstanding in the presented threat classification, as well as, in provided body
of evidence, for real attacks on space infrastructure, is the fact, that ground
stations along with the mission control center(s), are critical choke points
of the trust in, and reliance on, satellite systems correct operation.

Indeed, many malicious activities are enabled if attacks can be conducted
through the ground segment. Of course, countermeasures are available, in
particular, air-gapping ground stations and mission control. Such measures can
only make work of attackers a bit more challenging, but, will not prevent them
from setting up their own rogue infrastructure or creating of targeted malware
aimed at ground and space segment, thrown over the air-gap.

The deep reconnaissance of ground-and spacebased operational technology, will be
the basis of all the attacks. To mount space system attacks, adversaries have to
start by acquiring knowledge about the hardware/software stack of the
satellite. This knowledge may be obtained from the bill-of-materials, leaked
system specifications, or theft of FMECA documentation.  Careful analysis of
open hardware, open software and documentation collected from NewSpace vendors
will help building up the required understanding of the space systems'
capabilities, its limits and intended way of operation. Investigation and
review of software components will be useful, especially if open source code is
used on-board or on ground systems of interest. Obtained code can be analyzed
for vulnerabilities or can be used for opening the supply chain attack vector.
At this stage it will be of great importance to find out about safe-modes and
contingency procedures of investigated systems. Attackers shall be interested
in looking for ways of inducing the safe-mode, to either affect the
functionality of a system (in safe mode system ensuring their own survival will
very likely less secure than in nominal modes) or to conceal the breach.

The 2nd tier of cyberphysical attack requires obtaining access to a ground
station, either by establishing a rogue station or by gaining an unauthorized
access to an existing one.
The former requires perpetrators to have valid cryptographic keys, full
knowledge of the protocol stack and equipment capable of acquiring and
maintaining radio links to the satellite, while the latter requires them to
have access to the ground facility and remain undetected for long enough to
complete their attack. With the advent of heavily distributed, Ground Station
as a Service ventures this path becomes easier, as the more stakeholders mean
more chances of some party not conforming to security regulations. On the other
hands, similar effect could be achieved by tinkering with user terminals,
especially if are mass produced for Internet of Things, but many examples show
that industrial grade VSAT can be easily compromised and provide unauthorized
access to services.

Accessing the 3rd tier of cybyr-physical attack on space system involves
tracking the space system and monitoring it's behavior.
Tracking reveals when the satellite is in the communication windows
with the compromised (or rogue) ground station and for how long this window will
remain open. Information about the length of this visibility window is
crucial for select short slots (that might not be used by operators)
for test connections or to select the ones that are sufficiently long
to finalize the attack on the space system in one attempt.
Monitoring the system behavior requires reading out telemetry reports and
keeping track of system power input, battery charge state, current loads, mode
of operation, payload status, reaction wheel speeds, is required to know what
type of attack is suitable and when the attack is most effective to damage the
vehicle(e.g. by draining the batteries by switching on loads while in eclipse,
changing attitude to burn optical payload by directing it to Sun, attempting to
change the orbit by enabling the propulsion, switching to safe mode, if it is
not safe to do so).

By satisfying the requirements of first three stages, adversaries will
be well suited for conducing their attack successfully, taking control
over the space vehicle and inflicting damage to the system, or by
leading it to destroy itself or others.\\

One inherent weakness in today's operation of the space segment is
that vehicles in the latter accept commands and tasks received from
ground as fully trusted as long as they can decode these commands
correctly. Sanity checks and more elaborate communication schemes that
would prevent total system takeover in case a ground facility is
compromised are rarely considered or applied.

Moreover, with the proliferation of standard hard- and software
components, targeted exploits not only affect single spacecraft, but
may bring down whole constellations of satellites. We already see this
effect on ground based systems where homogeneity of automotive
equipment bears a similar threat of adversaries taking ov`er entire
fleets. In all that, satellites must be considered not only as digital
assets but as cyber-physical systems with all involved threats, both
to the satellite itself and to its environment if it gets under
control of an adversary~\cite{Oakley20, Mazzolin20}.

While attacks on the user segment are "low hanging fruits" that can be
relatively easily reaped benefit from,
%
%
attacks on space segment require a more coordinated effort and
substantial resources.


Figure~\ref{fig:space-cps-attack-pyramid}
shows the subsequent stages on which an attacker has to focus to mount
his/her space cyber-physical system attack.



\section{What Must Change?}
\label{sec:countermeasures}
\st{Here we discuss and point to main areas which offer improvements}

Certainly, we should not ignore the possibility of future targeted
cyber attacks to space vehicles. However, unfortunately the solutions
we apply to protect ICT systems on the ground, primarily following the
arms race between new threats being revealed and patches trying to fix
known vulnerabilities, only apply to a limited extent to space
vehicles (and they have not be proven very effective in the past). We
have to try to anticipate attacks, assume they will be partially
successful and prepare the system for tolerating such attempts without
already causing damage in order to buy the time for other mechanisms
to return the space vehicle to a state at least as secure as before
the attack.

In light of threats yet to reveal themselves over the multi-decade lifespan of
many satellites, coupled with the virtually non-existent (as of the moment of
writing) physical access and hardware upgrade capabilities to the satellites,
once deployed, it will be essential to prepare upfront for any form of recovery
that may have to be applied later.

\subsection{Resources}

Essentially the above strategy boils down to ensuring that sufficient resources
will be available at all points in time to throw out adversaries or compensate
for occurrence of complex, sometimes Byzantine faults,  and return the system
to security and, most importantly, safety for both, the vehicle and its
environment. With care, these resources will be exclusively computation
resources, needed to improve the software stack over time, but also to retain
the degree of replication that is already available to ensure safety despite
accidental faults.

The critical element of the above sentence is availability, in particular in the
presence of targeted attacks. Loosing static trust anchors (e.g., because of
compromised ciphers) will grant adversaries full control over the system
(assuming that the telecommand channel has already been taken over by
adversaries). On the other hand, when considering reconfigurable trust anchors,
the same reconfiguration interface, required to build long-term security by
continuously updating said trust anchor, presents itself, as an attractive
opportunity for adversaries. This opportunity not only widens the satellites
attack surface, but also gives them the opportunity to install a permanent
foothold. 

In the following, we review some of the resources of a space vehicle
to derive guidance for future system architectures and their
assurance processes.

\subsection{Platform and Payload}


As provided body of evidence shows, both platform and payload can be attacked in
cyber and physical domains. They suffer from single point of failure syndromes,
which despite deploying extensively redundant architectures nowadays, are
inevitable (i.e. especially payloads are not expected be fully redundant).
Today, the only known principle measures to mitigate single point of failure
syndromes  are the construction of components that cannot fail (a futile
endeavor in highly radioactive environments and with systems of not trivial
complexity) and the replication and distribution of functionality, so that not
all replicas fail simultaneously. Replication may and shall happen at many
levels: starting from components constituting the system, to the systems
(crafts, vehicles) up to the system-of-systems (swarms, formations,
constellations). However, replication is necessary but not sufficient to
achieve the goal of unconditional fault mitigation. If the replicas do not
cooperate actively to tolerate the failures or compromises, the dedicated
attacker or unfortunate accidental fault will overcome the protections and
posses the system. Hence, to further eliminate the residual risk of compromise
and, in particular, of fault propagation, outcomes of replicated system shall
be applied only after consensus on the outcome, has been reached. Consensus
becomes the last line of defense (e.g., after plausibility checks have
confirmed the validity of control signals), especially when interfacing to
actuators or while accepting commands and appending the on-board schedule.

All that being said about replication, in space context, it is not an absolute
remedy for all safety and security concerns, mainly due to power, mass,
bandwidth and accessibility constraints that the space infrastructure exhibits.
Therefore, we advocate further research in ways of accommodating consensus
based replication, especially taking into account incoming technology
improvements (efficient resource utilization) and proliferation of distributed
space systems.




\subsection{Ciphers}

Ciphers securing data, but more importantly command streams, including the
software or gateware updates and fixes are and will be required, we therefore
have to anticipate that the cipher, its implementation or the used keys need
update, possibly much more frequent than the remaining software stack itself.
Some of these elements can be constructed from others (like session keys from a
secret possessed and used to authenticate a node, and the latter being derived
from a host key to limit how often this host key is exposed).
However, the root algorithm itself and the key used to receive it in case all
other levels are compromised, remain critical.
To also replace this root of trust, sufficient resources must be
provided to host future replacements of the root encryption algorithm,
including sufficient memory to hold the new root key. Then the root
key can be used to decrypt the received replacement, before the
vulnerability can be exploited, if necessary proactively, in case the
security of the root algorithm is at risk.

From this point onward, patches can be validated and installed, and
the subsequent command sequence authenticated and applied.

\subsection{Communication}

%

Communication links, both ground and space ends, are critical points of entry
that deserve special attention. First, those subsystems are complex, thus
inherently contain exploitable vulnerabilities and are exposed directly to adversaries, in such
way we can't prevent their attempts to tinker with defended infrastructure.

In our opinion, the only viable solution can be found in replication mechanism
extending into time and space domains. In time domain, it is outpacing
adversaries in compromising the critical communication links needed to patch
subsystems themselves if this enable bypassing the authentication and
authorization mechanisms.. This implies frequent resets of these most critical
communication systems, possibly in combination with replication to circle
through just repaired entities, which the adversary would need to compromise
again before she can continue with its attack. In space domain, replication
means having mutually independent commanding paths (direct but also relayed
through other networks and nodes), commanding centers (coordinating, among
themselves, on the ground, the commanding actions and schedules) and on-board
decoders capable of sharing collected inputs form mission controls and
exercising the consensus algorithms on them. While such approach would improve
the space infrastructure safety and security enormously, as of the moment of
writing, it has not been implemented.

\subsection{Processes and Assurance}

Of course, for all elements where speed is of essence, in particular
fixing vulnerabilities, existing validation and assurance processes
are simply too slow to outpace adversaries in their doing. Pre-assured
components may be a solution, but absolute confidentiality needs to be
applied to ensure knowledge on how to attack them is not leaked to the
adversary. A possibly better option would be to restrain components to
the roles they have to play and live with the residual risk of patches
introducing new vulnerabilities, which adversaries would still need to
identify before they can be exploited. Then, with enough time outside
the critical moments of the satellite being under attack, patches can
be hardened to improve their correctness.

Restraining requires limiting access to resources to only those
essential for the purpose of a component and may, as described for
actuators as well benefit from replication and voted access to all
configuration possibilities.


\section{Conclusions}
\label{sec:conclusions}
\st{finishing thoughts}
In this paper, we have surveyed existing and reported attacks to
develop a comprehensive survey of threats space systems are exposed
to. Unlike previous studies, the attacks we are most worried about
include cyber attacks possibly mounted by small groups of hackers with
relatively simple equipment or leveraging the compromised equipment of
space-faring nations to take control over space vehicles with possibly
severe consequences on the defense capabilities, but more importantly
the economy of developed countries. Adversaries compromising GNSS may
impact navigation, logistics and other businesses that depend on this
service, and with satellite control in the hands of such groups, they
may even turn the captured vehicle into a cyber-kinetic weapon,
targeting other space craft or ultimately triggering Kessler's
syndrome, which would render space inaccessible to developed nations
for the upcoming decades until debris capture technology becomes
available.

In our opinion, such asymmetric threats are best countered by leveraging on the
already widely deployed accidental fault tolerance mechanisms to prepare
satellites to also tolerate targeted attacks to buy the time required to
rejuvenate them to a state at least as secure as initially. Clearly, such an
effort requires significant changes, including of the assurance processed
deployed for ``Old Space'', but more importantly also for ``New Space''
equipment, but also increased situational awareness and the will and resources
(partially in space) to achieve this tolerance. 

We can predict fault and mitigate risks accordingly, but some faults, even if
obeying statistics might be Byzantine for which the systems might not be ready.
Byzantine fault is accidental equivalent of malicious activities. It might be
interesting not to focus on safety only, or on security only, but to focus on
Byzantine fault models instead and building the systems with Byzantine fault
tolerance.

In context of this paper it is interesting exercise to read about vision of
futuristic counterspace activities as written by Zielinski et al. in the
90's~\cite{Zielinski96} for 20th century. What's striking during this lecture is
that many of the predictions became reality much sooner than anticipated
(extreme miniaturization, satellite cloaking, ground based high energy lasers,
proximity operations and satellite bodyguards, precise optical and radar
tracking)  later became a reality with the rest of the concepts quickly
catching up. We believe, the same will happen to cyber-physical warfare in
space - once implausible, unlikely, too expensive, not practical, in near
future will become everyday reality of space exploration and exploitation.

\bibliographystyle{IEEEtran}
\bibliography{space-threat-vectors}

\end{document}